\documentclass{aa}

\usepackage[varg]{txfonts}
\usepackage{graphicx}

\usepackage{booktabs}
\usepackage{tabularx}
\usepackage{multicol}
\usepackage{subcaption}

\begin{document}

\title{Cosmic insights from galaxy clusters: Exploring magnification bias on sub-millimetre galaxies}
\titlerunning{Cosmic insights from galaxy clusters}
\authorrunning{Fern{\'a}ndez-Fern{\'a}ndez R. et al.}

   \author{Fern{\'a}ndez-Fern{\'a}ndez R.\inst{1,2} \and Bonavera L.\inst{1,2} \and Crespo D.\inst{1,2} \and Gonz{\'a}lez-Nuevo J.\inst{1,2} \and Cueli M. M.\inst{3,4} \and Casas J. M.\inst{1,2} \and Cabo S. R.\inst{1,2} }

  \institute{
  $^1$Departamento de F\'{i}sica, Universidad de Oviedo, C. Federico Garc\'{i}a Lorca 18, 33007 Oviedo, Spain\\
    $^2$Instituto Universitario de Ciencias y Tecnolog\'{i}as Espaciales de Asturias (ICTEA), C. Independencia 13, 33004 Oviedo, Spain\\
  $^3$SISSA, Via Bonomea 265, 34136 Trieste, Italy\\
    $^4$IFPU - Institute for fundamental physics of the Universe, Via Beirut 2, 34014 Trieste, Italy\\         
}
\date{}
\abstract
{Magnification bias, an observational effect of gravitational lensing in the weak regime, allows the cosmological model to be tested through angular correlations of sources at different redshifts. This effect has been observed in various contexts, particularly with sub-millimetre galaxies (SMGs), offering valuable astrophysical and cosmological insights. 
}
{The study aims to investigate the magnification bias effect exerted by galaxy clusters on SMGs and its implications for astrophysical and cosmological parameters within the $\Lambda$ cold dark matter model.
}
{Magnification bias was explored by quantifying the cross-correlation function,which we then utilised to derive constraints on cosmological and astrophysical parameters with a Markov chain Monte Carlo algorithm. Two distinct galaxy cluster samples were used to assess result robustness and understand the influence of sample characteristics. 
}
{Cluster samples show higher cross-correlation values than galaxies, with an excess at larger scales suggesting contributions from additional large-scale structures. The parameters obtained, while consistent with those of galaxies, are less constrained due to broader redshift distributions and limited cluster statistics. Results align with weak lensing studies, hinting at slightly lower $\sigma_8$ and $\Omega_m$ values than \emph{Planck}'s cosmic microwave background data, emphasising the need for enhanced precision and alternative low-redshift universe tests.
}
{While this method yields constraints that are compatible with the $\Lambda$ cold dark matter model, its limitations include broader redshift distributions and a limited number of lenses, resulting in less constrained parameters compared to previous galaxy studies. Nonetheless, our study underscores the potential of using galaxy clusters as lenses for magnification bias studies, capitalising on their elevated masses and thus providing a promising avenue to test current cosmology theories. Further progress can be made by expanding the lens sample size.
}


\maketitle

\section{Introduction}
Over the past few decades, gravitational lensing has emerged as a versatile tool in observational cosmology. Essentially, the distribution of matter -- both baryonic and dark -- across the universe, including galaxies and clusters of galaxies, exerts a bending effect on light as it travels through space. This phenomenon leads to the apparent amplification or deformation of the brightness and size of distant objects \citep[e.g.][]{SCH92}.

Among the observational effects of gravitational lensing in the weak regime, a notable phenomenon is the statistical surplus of high-redshift sources in the vicinity of low-redshift high-mass structures, known as magnification bias \citep[a comprehensive review can be found in][]{Bar01}. The magnification bias effect manifests itself as a discernible angular cross-correlation function between two sets of sources that span different redshift ranges, indicating that the large-scale structure traced by the foreground sources is amplifying the background ones. Since the gravitational deflection of light travelling close to the lenses depends on cosmological distances and galaxy halo properties, magnification bias can be used as an independent cosmological probe to estimate parameters in the standard cosmological model, complementing other cosmological probes.

Magnification bias has been observed across various contexts, including the correlation between low-redshift galaxies and high-redshift quasars \citep[e.g.][]{SCR05, MEN10}, \textit{Herschel} sources and Lyman-break galaxies \citep[e.g.][]{HIL13}, and even between the cosmic microwave background (CMB) and other source distributions \citep[e.g.][]{BIA15, BIA16}. Among the array of background sources, sub-millimetre galaxies (SMGs), characterised by their steep luminosity function, high redshifts, and faint optical band emission, stand out. These properties have been well quantified, and SMGs can be considered solid targets for study through lensing \citep{GON12, BLA96, Neg07, NEG10, NEG17, GON17, BUS12, BUS13, FU12, WAR13, CAL14, NAY16, Bak20}. Previous studies have also revealed a strong magnification bias effect in SMGs, resulting in measurements that unveil valuable astrophysical and cosmological insights \citep[][]{GON17, BON20, CUE21, GON21, Cue23, BON23}.

Furthermore, dividing the foreground sample into distinct redshift bins has been considered. The findings suggest that this approach paves the way for a more thorough tomographic analysis \citep[][]{BON21, CUE22}, thereby enhancing the precision of the obtained astrophysical and cosmological constraints.

This study analyses, for the first time, the magnification bias generated by galaxy clusters acting as lenses on SMGs. Our investigation focused on measuring the cross-correlation function, which we subsequently leveraged to derive constraints on both cosmological and astrophysical parameters through the use of a Markov chain Monte Carlo (MCMC) algorithm. Two distinct galaxy cluster samples were employed to test the robustness of the results and assess the influence of the samples' main characteristics.

This article is organised into the following sections: In Section \ref{sec:method} an overview of the background sources and lenses used is provided along with a concise summary of the methodology adopted in this study, which encompasses measurements and parameter estimations. Section \ref{sec:results} presents the measurements and results from the analysis. Finally, Section \ref{sec:concl} presents the key conclusions drawn from this investigation. The most relevant aspects of the theoretical framework related to the weak-lensing-induced magnification bias are outlined in Appendix \ref{app:framework}. Appendix \ref{app:corner plots} presents corner plots showing the marginalised posterior distribution and probability contours of the parameters estimated in all the cases presented in this work. Finally, Appendix \ref{app:xcorr_zones} provides the analysis of the cross-correlation function for each Galaxy And Mass Assembly (GAMA) zone separately.
\begin{figure}[ht]
\includegraphics[width=0.5\textwidth]{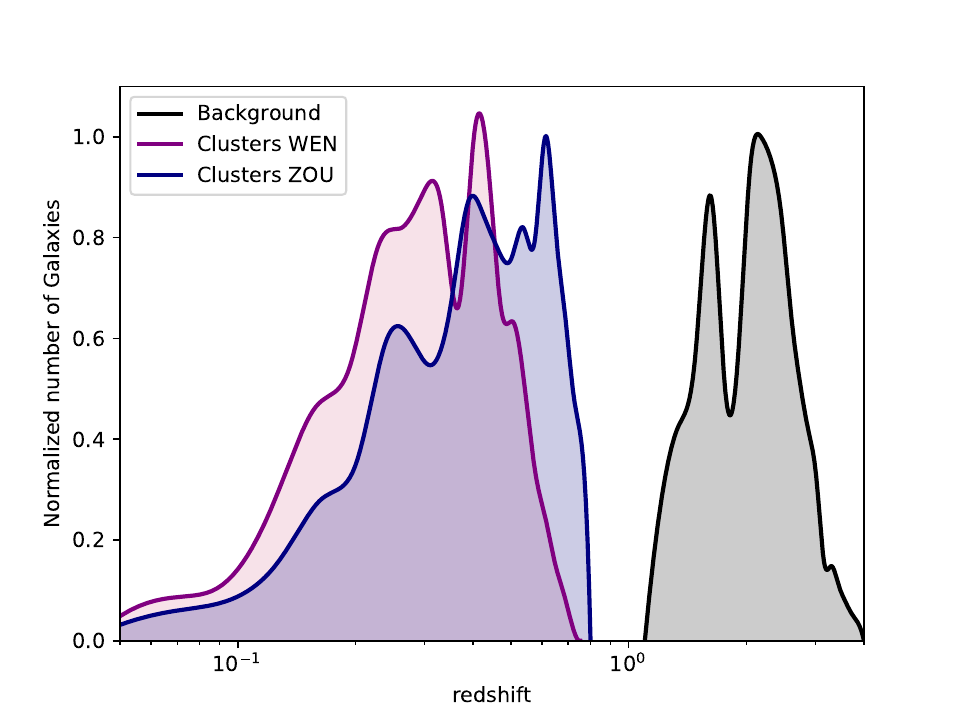}
 \caption{Redshift distributions of the background SMGs (in black) and foreground lenses from ZOU (blue) and WEN (purple) clusters.}
 \label{fig:z_distro}
\end{figure}

\section{Methodology}
\label{sec:method}
\subsection{Data}
\label{subsec:data}

In this study we employed two distinct catalogues of galaxy clusters as lens samples to investigate the cross-correlation induced by weak lensing, using a set of high-redshift SMGs as background sources. Additionally, we included cross-correlation measurements from a previous study conducted by \citet{Cue23}, who utilised a sample of galaxies as lenses, for comparative purposes. Further details regarding the data employed for calculating the cross-correlation signal are comprehensively discussed in \citet{GON23} and \citet{CRE24}.

Specifically, the background sample was selected from the \textit{Herschel} Astrophysical Terahertz Large Area Survey \citep[H-ATLAS;][]{EAL10} sources within the GAMA fields (G09, G12, and G15) and the north and south galactic poles (NGP and SGP). The GAMA fields are located at celestial equatorial positions of 9, 12, and 14.5 hours, covering an approximate common area of $\sim147$ deg$^2$, in addition to the NGP and SGP areas, covering an additional 180.1 and 317.6 deg$^2$. To avoid redshift overlap with the lenses, a photometric redshift selection of $1.2<z<4.0$ was applied. The photometric redshifts were estimated using a $\chi^2$ fitting procedure of the photometric data to the spectral energy distribution of SMM J2135-0102 \citep{LAP11,GON12, IVI16}. The resulting SMG dataset comprises around 66,000 sources with an average photometric redshift of 2.20. Their redshift distribution, estimated as $p(z|W)$ considering galaxies selected by a window function within the redshift range of $1.2<z<4.0$, is shown in black in Fig.\ref{fig:z_distro}. One of the lens samples did not cover the SGP region, prompting its exclusion from the analysis. Considering only the utilised area, we were left with 56874 sources in the background sample.

The lens samples were obtained from the \citet[][]{ZOU21} catalogues (hereafter, ZOU) and \citet[][]{WEN12}, catalogues (WEN), both of which originated from the Sloan Digital Sky Survey (SDSS). The ZOU cluster catalogue comprises 540432 clusters, covering approximately 20000 $\text{deg}^2$ at redshifts $z < 1$ within the DESI (Dark Energy Spectroscopic Instrument) legacy imaging surveys \citep{DEY19}. These clusters have a median redshift of $0.53$, a median richness of 22.5 (with a lower limit of 10), and a mean mass of $1.23\times10^{14} M_{\odot}$. The selected sample overlapped with the H-ATLAS zones G09, G12, G15, and NGP, and it is restricted to $z<0.8$ to differentiate it from the background galaxies. The final selection consisted of 9,056 clusters, with a redshift distribution shown in Fig. \ref{fig:z_distro}. The mean redshift of the final sample is $\langle z\rangle=0.50^{+0.24}_{-0.30}$; the errors were determined considering the 95\% confidence intervals derived from the redshift distribution. It should be noted that the available cluster data in this catalogue offer two different sets of possible central positions for each object, based on the brightest cluster galaxy (BCG) or based on the mass peak (PEAK hereafter). Throughout this work, BCG data were used unless explicitly stated otherwise.

The WEN cluster catalogue comprises 132684 clusters with redshift $0.05\leq z < 0.8$ and a median richness of 17 (with a lower limit of 12), identified using the photometric redshifts of galaxies from SDSS-III. After selecting sources that overlap with the three GAMA fields and the NGP region, a final sample of 3598 galaxy clusters was obtained, with a mean redshift of $\langle z\rangle=0.38^{+0.23}_{-0.22}$. The redshift distribution is shown in Fig. \ref{fig:z_distro}. Table \ref{tab:clusters} summarises the lens samples characteristics.

\begin{figure}[ht]
    \centering
    \includegraphics[width=0.5\textwidth]{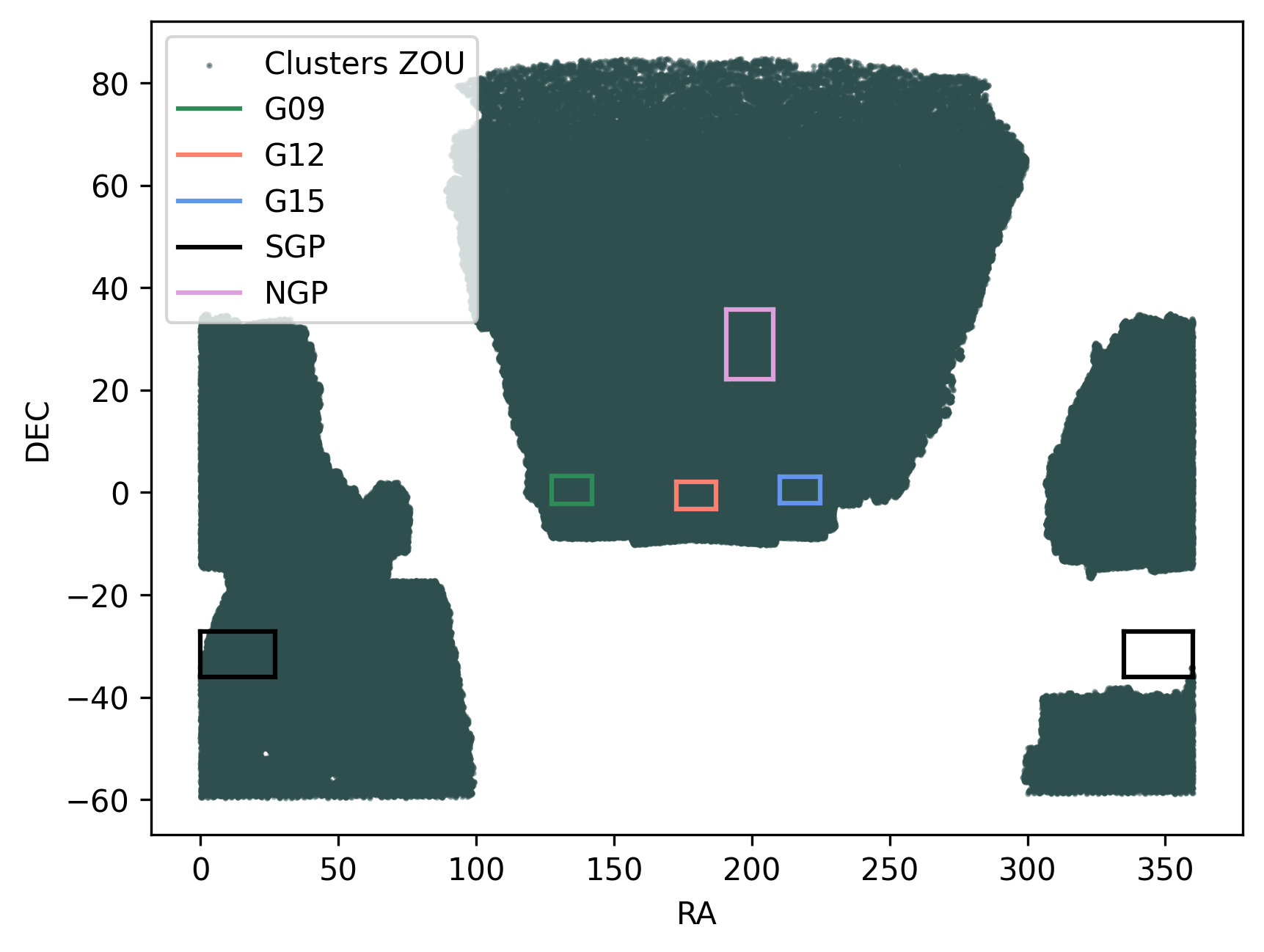}
    \includegraphics[width=0.5\textwidth]{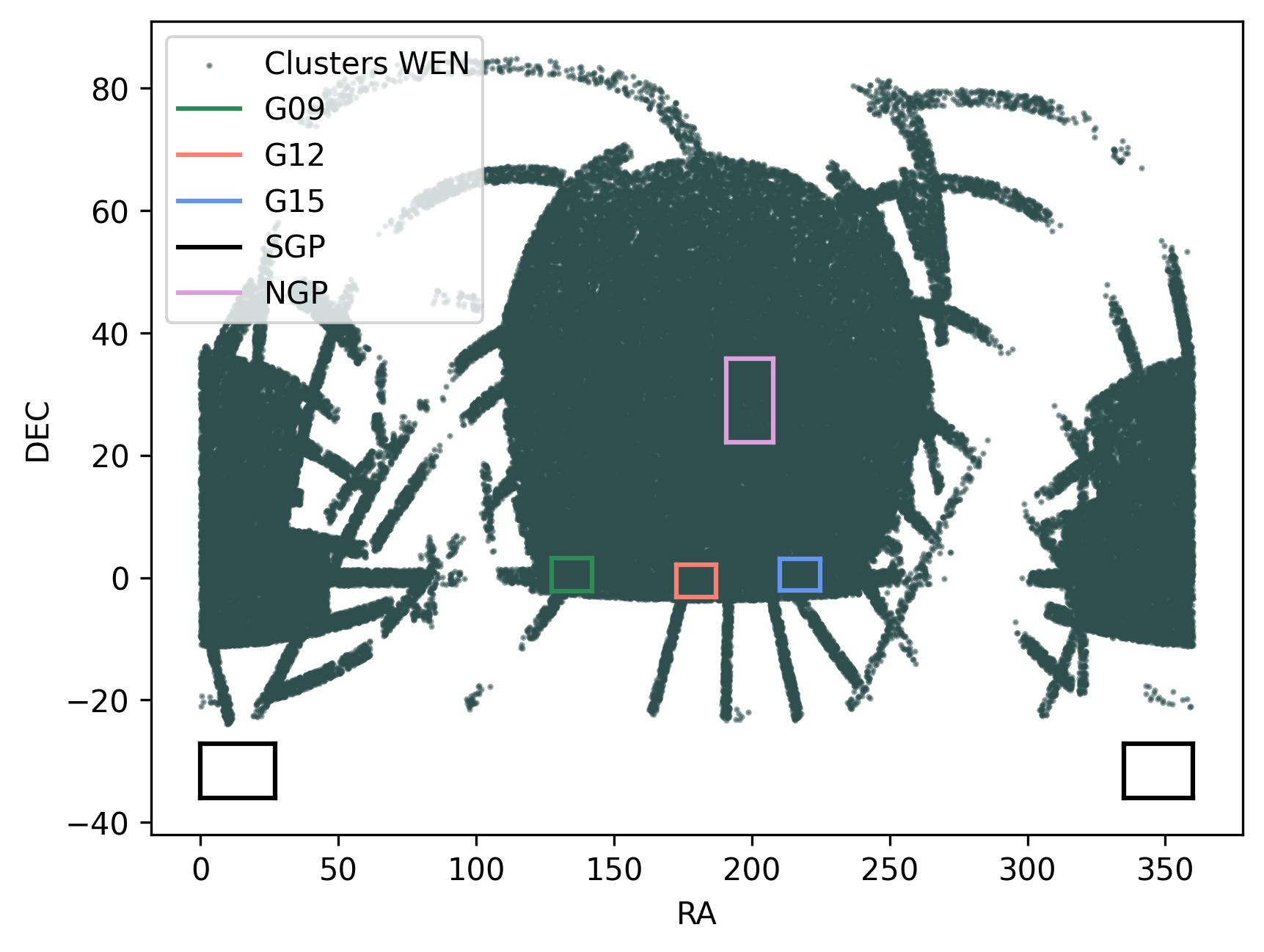}
    \caption{Lens catalogue RA-Dec. distributions for ZOU clusters (upper panel) and WEN clusters (lower panel). In both cases, the position is given in terms of the BCG. The Gamma and NGP zones are delimited in both panels.}
    \label{fig:clustersRADEC}
\end{figure}

Finally, Fig. \ref{fig:clustersRADEC} displays the RA-Dec. distribution of the foreground lens samples based on the BCG positions: ZOU in the upper panel and WEN in the lower one. The GAMA, SGP, and NGP zones are also shown to account for the real area used in this work. The SGP area was not considered this time, although, as shown in the upper panel, around 50\% of it could be used to further improve the statistics of ZOU cluster sample.

\begin{table}[ht]
  \caption{Summary of the lens samples.}
  \label{tab:clusters}
  \centering
  \small
  \begin{tabularx}{\columnwidth}{ccccccc} 
    
    \hline\hline
    Sample & Sources & $\langle z \rangle$  & 
    G09 & G12 & G15 & NGP\\
    \midrule
    ZOU & 9056 & 0.50 &
    1490 & 1412 & 1550 & 4604 \\
    WEN & 3598 & 0.38 & 
    574 & 588 & 613 & 1823 \\
    \hline\hline
  \end{tabularx}
  \tablefoot{The first column is the name of the catalogue (sources are clusters, unless stated otherwise), the second column represents the total number of sources, the third the average redshift.  Subsequent columns detail lens counts across designated regions (GAMA and NGP).}
\end{table}


\subsection{Measurements}
\label{subsec:measurements}
The cross-correlation function and covariance matrix were measured according to the method thoroughly discussed in \cite{GON23} and summarised in \cite{Cue23}. Briefly, this approach combines the counts of distinct foreground-background pairs from each field, resulting in a single estimation. To compute the covariance matrix function the complete available area (GAMA zones and NGP) is subdivided into smaller sections employing a bootstrap approach that capitalises on the entire field area.The cross-correlation function is then measured using the modified version of the \cite{LAN93} estimator by \cite{HER01},

\begin{equation}
    \label{eq:w_fb_obs}
    \hat{w}_{\text{fb}}(\theta)=\frac{\rm{D}_f\rm{D}_b(\theta)-\rm{D}_f\rm{R}_b(\theta)-\rm{D}_b\rm{R}_f(\theta)+\rm{R}_f\rm{R}_b(\theta)}{\rm{R}_f\rm{R}_b(\theta)},
\end{equation}where $\rm{D}_f\rm{D}_b$, $\rm{D}_f\rm{R}_b$, $\rm{D}_b\rm{R}_f$, and $\rm{R}_f\rm{R}_b$ represent the normalised foreground-background, foreground-random, background-random and random-random pair counts at an angular separation $\theta$. For the generation of the background random samples the influence of the \textit{Herschel} scanning strategy on the surface density variation was accounted for.

Then, the covariance matrix was computed via the bootstrap technique, which is based in randomly re-sampling the data to obtain a number of sub-samples of a given size. More specifically, Bootstrapping implies dividing the whole common area into $N$ equal-surface area sub-regions, resampling $N_r$ of them with replacement and then repeating the process $N_b$ times. As explained in \cite{GON23}, the subsamples or patches are obtained using a k-mean algorithm, which is designed to maintain the spatial dependence structure intact during the re-sampling procedure. This new methodology has demonstrated enhanced robustness and significantly reduced uncertainties in the obtained measurements. For this work, the entire field was partitioned into a minimum of $N=20$ sub-regions or patches to guarantee an ample count of measurements. A total of $N_b=10000$ bootstrap samples were generated using an oversampling factor of 3 (that is, $N_r=3N$), which, as suggested by \cite{NOR09} and validated in \cite{GON23}, contributes to a more reliable analysis. The covariance matrix is given by

\begin{equation}
    \text{Cov}(\theta_i,\theta_j)=\frac{1}{N_b-1}\sum_{k=1}^{N_b}\,\bigg[\hat{w}_k(\theta_i)-\bar{\hat{w}}(\theta_i)\bigg]\bigg[\hat{w}_k(\theta_j)-\bar{\hat{w}}(\theta_j)\bigg]\label{covariance},
\end{equation}where $\hat{w}_k$ is the measured cross-correlation function from the $k^{\text{th}}$ Bootstrap sample and $\bar{\hat{w}}$ is the corresponding average value over all Bootstrap samples.

\begin{figure}[ht]
\includegraphics[width=0.5\textwidth]{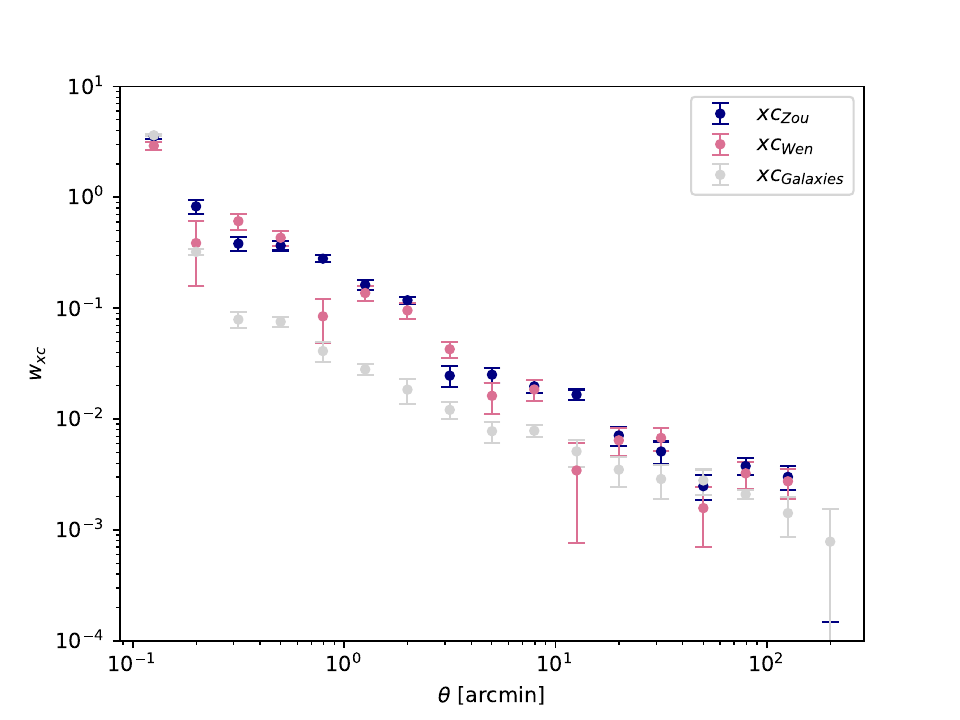}
 \caption{Cross-correlation data obtained with ZOU (blue circles) and WEN (pink circles) clusters as lenses. For comparison, cross-correlation data obtained with the galaxy sample by \cite{Cue23} are shown in grey.
 }
 \label{fig:xc_data}
\end{figure}

Figure \ref{fig:xc_data} shows the measured cross-correlation data obtained with both cluster lens samples. As a comparison we  also show the measurements obtained in \cite{Cue23} using a lens sample of approximately 150000 galaxies with 0.2 < $z_{spec}$ < 0.8 from the GAMA II spectroscopic survey \citep{DRI11,BAL10,BAL14,LIS15}. This is the same galaxy sample used as lenses in previous studies \citep[][]{GON17, BON20, CUE21, GON21, GON23, Cue23, BON23}.

The cross-correlation values are noticeably stronger for both galaxy clusters in comparison to the galaxy population. This trend is particularly prominent at low and intermediate angular distances and can be attributed to the different mass scales associated with each lens sample, a topic that will be examined in detail in Section \ref{sec:results}. Furthermore, the ZOU data transition from the one-halo to the two-halo regime, occurring around $\sim 2-3$ arcmin (equivalent to $\sim 1$ Mpc), displays a more pronounced abruptness compared to the measurements obtained from the galaxy sample.

Remarkably, both cluster datasets exhibit an unexpected `bump' in their cross-correlation measurements, notably apparent at angular scales exceeding 60 arcmin. Intriguingly, a similar phenomenon was previously documented in studies involving galaxy samples \citep{GON23, Cue23}. In the case of \citep{Cue23}, the excess signal observed at large scales was attributed to sampling variance. However, the recurrence of this distinctive feature in the cluster data, despite possessing distinct systematic characteristics, prompts us to consider a potential common physical origin.

Since the selected GAMA zones are consistent with those used in \citet{Cue23}, we conducted a parallel analysis for both the WEN and ZOU catalogues studying the cross-correlation function for each individual zone. Detailed results are available in Figures \ref{fig:zou_zones} and \ref{fig:wen_zones}, provided in Appendix \ref{app:xcorr_zones}. Notably, the cluster datasets also exhibit pronounced sample variance at large angular scales. The excess signal is primarily driven by the G15 contribution in both cluster catalogues, with a supplementary positive contribution from G12. Additional evidence supporting this signal excess is also presented in Appendix \ref{app:xcorr_zones}. In Figure \ref{fig:npairs}, normalised foreground-background pair counts (the $\rm{D}_f\rm{D}_b(\theta)$ term in Eq. \ref{eq:w_fb_obs}) are depicted for both cluster samples within each specific zone. They are normalised to the random alignment pair counts estimated from mock
foreground and background random catalogues (the $\rm{R}_f\rm{R}_b(\theta)$ term in Eq. \ref{eq:w_fb_obs}). The excess at small angular scales is simply an indication of a non-negligible cross-correlation signal. On the contrary, at the largest angular scales, the data-data pair counts are expected to converge to the random ones. However, it is noteworthy that both G12 and particularly G15 exhibit a noticeable increase beyond 60 arcminutes that cannot be attributed to cluster lensing. These findings hint at the possibility of additional large-scale structures, such as superclusters or filaments, contributing to the observed signal. 

\begin{figure}[ht]
\includegraphics[width=0.5\textwidth]{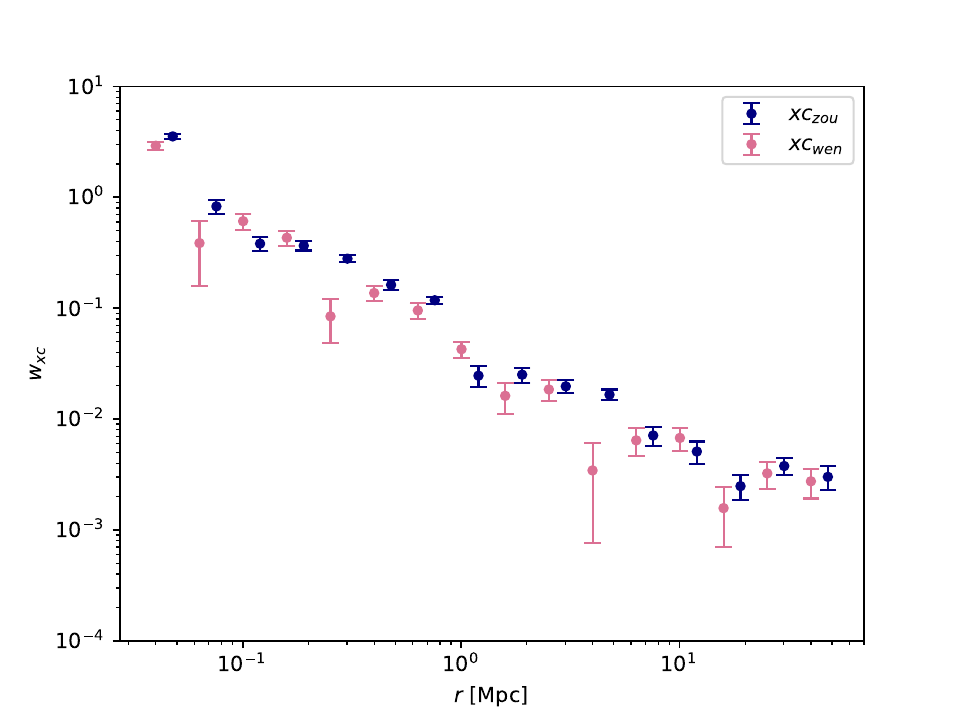}
 \caption{Cross-correlation dependence on physical distance for the ZOU cluster (in blue) and the WEN cluster (in pink). The distance was calculated using the mean redshift values for each cluster, specifically $z=0.5071$ for ZOU and $z=0.3724$ for WEN.}
 \label{fig:xc_mpc}
\end{figure}

Given that the two cluster samples possess slightly different redshift distributions, the angular scales correspond to varying physical scales. In Fig. \ref{fig:xc_mpc}, we present a comparison of the cross-correlation measurements after accounting for the mean redshift of each sample. Notably, despite coming from unrelated catalogues, the two datasets exhibit a similar pattern. This congruence is evident in the behaviour observed in the transitional region from the one-halo to the two-halo regime, 1-3 Mpc, as well as the presence of the aforementioned anomalous bump at larger scales\footnote{For a theoretical overview of the cross-correlation and the HOD models under consideration, readers are directed to Appendix \ref{app:framework}.}, approximately exceeding 20 Mpc.

Another noteworthy aspect of the data pertains to the presence of outliers within the WEN dataset. These outliers manifest as depressions or dips in the cross-correlation, lacking a clear physical rationale and accompanied by elevated associated uncertainties. Given the substantial agreement between the cross-correlation results from both samples and the discernible differences in cluster counts (as seen in Table \ref{tab:clusters}), the most reasonable interpretation for these anomalous data points is a statistical representation issue. Therefore, it is reasonable to exclude them during subsequent analyses employing this dataset. 

Lastly, within the one-halo regime at small scales, a slight deviation in normalisation becomes discernible between the data obtained from both cluster samples. These disparities are expected to have implications for their respective halo occupation density (HOD) models, as elaborated upon in Section \ref{sec:results}, and could potentially correlate with the richness of each catalogue.

\subsection{Parameter estimation}
\label{subsec:paramest} 

In this study, the tested cross-correlation function model relies on a set of free parameters describing cosmology and the HOD model under examination. These parameters have been explored using the MCMC method, which employs a Bayesian statistical approach for inference. The purpose of this work is, therefore, to sample the posterior probability density function to obtain marginalised credible intervals for each of the parameters, as well as probability contours. The sampling was performed using the open-source software package {\texttt{emcee}}, developed by \cite{FOR13} and licensed under MIT. This package is built upon the affine invariant MCMC ensemble sampler introduced by \cite{GOO10} and is entirely implemented in Python. 

The measurement-based and theoretical cross-correlation functions are computed as described in Eqs. \ref{eq:w_fb_obs} (in Sect. \ref{subsec:measurements}) and \ref{eq:w_fb} (in Appendix \ref{app:framework}), respectively. The log-likelihood function is then obtained for $m$ number of measurements as a multivariate Gaussian in the form 

\begin{align*}
    \log{\mathcal{L}_{\text{cross}}\,(\theta_1,\ldots,\theta_m)}=-\frac{1}{2}&\bigg[m\log{(2\pi)}+\log{|C_{\text{cross}}|}\,+\\    &+\overrightarrow{\varepsilon}_{\text{cross}}^{\text{T}}C_{\text{cross}}^{-1}\,\overrightarrow{\varepsilon}_{\text{cross}}\bigg],
\end{align*}
where 
$\overrightarrow{\epsilon}_{\text{cross}}\equiv [\varepsilon_{\text{cross}}(\theta_1),\ldots,\varepsilon_{\text{cross}}(\theta_m)]$,

\begin{equation*}
    \varepsilon_{\text{cross}}(\theta_i)\equiv w_{\text{fb}}(\theta_i)-\hat{w}_{\text{cross}}(\theta_i)\quad\quad \forall i\in\{1,\ldots,m\},
\end{equation*}
and $C_{\text{cross}}$ is the covariance matrix of the cross-correlation measurements, evaluated as in Eq. \eqref{covariance}.

In this work, the number of walkers for each MCMC run was set to 21, three times the number of parameters to be estimated, and the number of iterations was set to 5000. In this procedure, $105000$ posterior samples were generated per run. These samples undergo reduction by discarding initial iterations and introducing a burn-in phase when flattening the chain. Each parameter's walkers were examined individually to identify the burn-in value. Iterations in which the walkers remained frozen or restricted to a single value or its vicinity, instead of exploring the parameter space, have been discarded. The number of iterations retained depended on the specific case and chains' convergence. 

Seven parameters were analysed in this study, which can be categorised into two groups: astrophysical and cosmological parameters. The astrophysical parameters are $M_{min}$, $M_1$, and $\alpha$. They define the HOD model, as described in the Appendix \ref{app:framework}.  The parameter $\beta$ (the logarithmic slope of the background sources’ number counts) has also been included following \citet{Cue23}. On the other hand, the cosmological parameters considered in this study include $\Omega_m$, $\sigma_8$, and $h$, where $h$ is defined as $H_0=100\cdot h\text{ km}\text{s}^{-1}\text{Mpc}^{-1}$. A flat universe is assumed, with $\Omega_{\text{DE}}=1-\Omega_m$. The values of $\Omega_b$ and $n_s$ are held fixed at their best-fit values from \citet{PLA18_VI}, which are $\Omega_b=0.0486$ and $n_s=0.9667$, respectively. Finally, in consideration of the weak lensing approximation, only the cross-correlation function data with angular scales greater than or equal to $0.5$ arcmin are taken into account, as thoroughly discussed in \citet{BON19}.

The choice of priors for each parameter is illustrated in Table \ref{tab:priors}. Uniform priors were adopted for both the astrophysical and cosmological parameters, with the latter being the same as those used in \cite{BON20, BON21}. Additionally, a Gaussian prior was applied to $\beta$, with the same values as in \citet{Cue23}.

\begin{table}[ht]
  \caption{Prior distributions for the MCMC analyses. }
  \label{tab:priors}
  \centering
  \begin{tabular}{cccc} 
    \hline\hline
    \multicolumn{2}{c}{Astro} & \multicolumn{2}{c}{Cosmo} \\
    \cmidrule(r){1-2} \cmidrule(r){3-4}
    Parameter & Prior & Parameter & Prior\\
    \midrule
    $\log{M_{min}}$ & $\mathcal{U}[12.5-15.5]$
        & $\Omega_m$ & $\mathcal{U}[0.1-0.8]$\\
    $\log{M_{1}}$ & $\mathcal{U}[12.6-15.6]$
        & $\sigma_8$ & $\mathcal{U}[0.6-1.2]$\\
     $\alpha$ & $\mathcal{U}[0.5-1.5]$
        & $h$ & $\mathcal{U}[0.5-1.0]$\\
     \midrule
     $\beta$ & $\mathcal{N}[2.8 , 0.1]$\\
     \hline\hline
  \end{tabular}
  \tablefoot{The first two columns provide details of priors for the astrophysical parameters, while the third and fourth columns summarise the cosmological priors. The prior for $\beta$ is presented separately below. Uniform priors (denoted as $\mathcal{U}$) are specified with minimum and maximum values for the selected interval. Gaussian priors ($\mathcal{N}$) are represented by their mean $\mu$ and standard deviation $\sigma$.}
\end{table}
\section{Results}
\label{sec:results}

In this section, the findings derived from our analysis of cross-correlation data within the context of this study will be presented. 
The results obtained from the ZOU and WEN catalogues individually will be discussed within its dedicated subsections. Particular emphasis will be placed on the ability of the HOD to serve as a tool for validating the outcomes obtained through MCMC fitting. To conclude this section, we undertake a comparative analysis, contrasting the findings from each catalogue with prior research centred on galactic studies.

\subsection{ZOU catalogue}
\label{subsec:zou}


As previously discussed in Section \ref{subsec:data} (and also depicted in Fig. \ref{fig:xc_data} or \ref{fig:npairs}), the cross-correlation data employing galaxy clusters as gravitational lenses, found in both the ZOU and WEN catalogues, deviates from the expected steep decline. Notably, it exhibits an increase at larger angular distances, with a pronounced effect observed at $\theta \gtrsim 60$ arcmin in the case of the ZOU catalogue. In light of the findings presented in \cite{Cue23}, who explored the impact of this large-scale anomaly on parameter estimation, we considered two distinct cases. The first case involved analysing the complete available range of angular scales above the prescribed lower limit. In the second case, measurements beyond $\gtrsim 60$ arcmin were excluded. Table \ref{tab:zou_results} summarises the values obtained for each parameter with both datasets. In each case, the first column shows the mean, the second the mode, and the third the 68\% credible interval (CI).

\begin{figure}[ht]
    \includegraphics[width=0.5\textwidth]{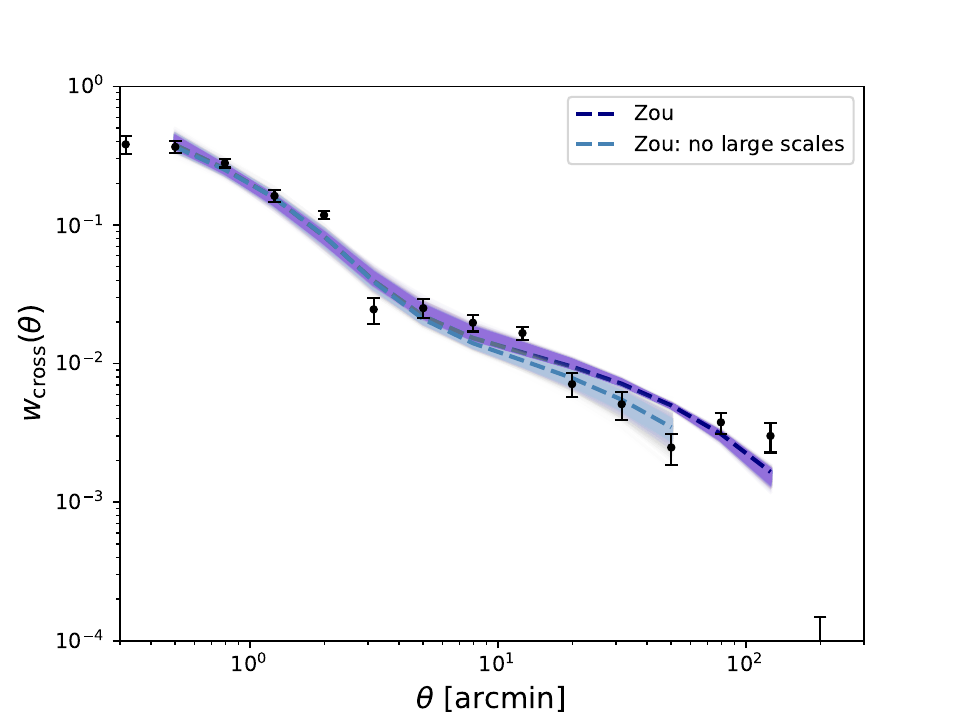}
     \caption{Sampling plots for ZOU. The dark blue and light blue regions illustrate the posterior sampling of the cross-correlation function derived from the entire dataset and the omitted large-scale cases (in which data exceeding $\gtrsim 60$ arcmin were excluded), respectively. Dashed lines indicate the best fit, and the data points are represented by black dots.
     }
     \label{fig:sampled_zou}
\end{figure}

\begin{figure*}[ht]
    \centering
    \includegraphics[width=0.32\textwidth]{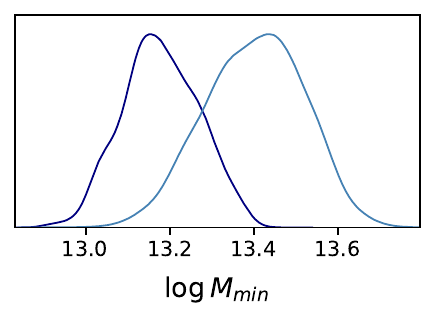}
    \includegraphics[width=0.32\textwidth]{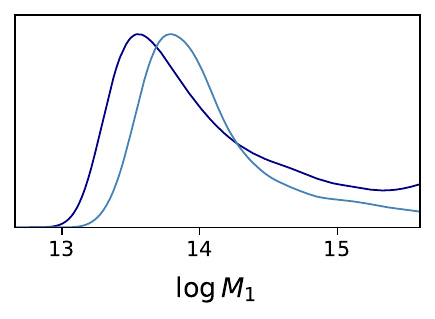}
    \includegraphics[width=0.32\textwidth]{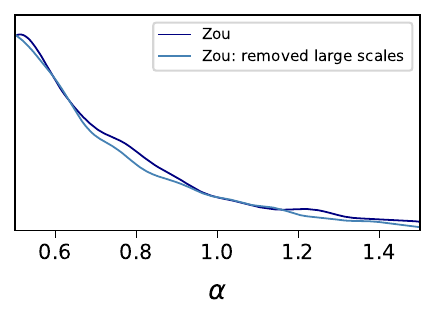}
    \caption{Marginalised posterior distribution for the HOD parameters obtained via MCMC runs on the cross-correlation function using ZOU cluster data.  The dark blue fit was obtained using the full available data, while  the light blue fit was obtained excluding cross-correlation data above $\gtrsim 60$ arcmin. 
    }
    \label{fig:zou_hod}
\end{figure*}

\begin{figure}[ht]
    \includegraphics[width=0.5\textwidth]{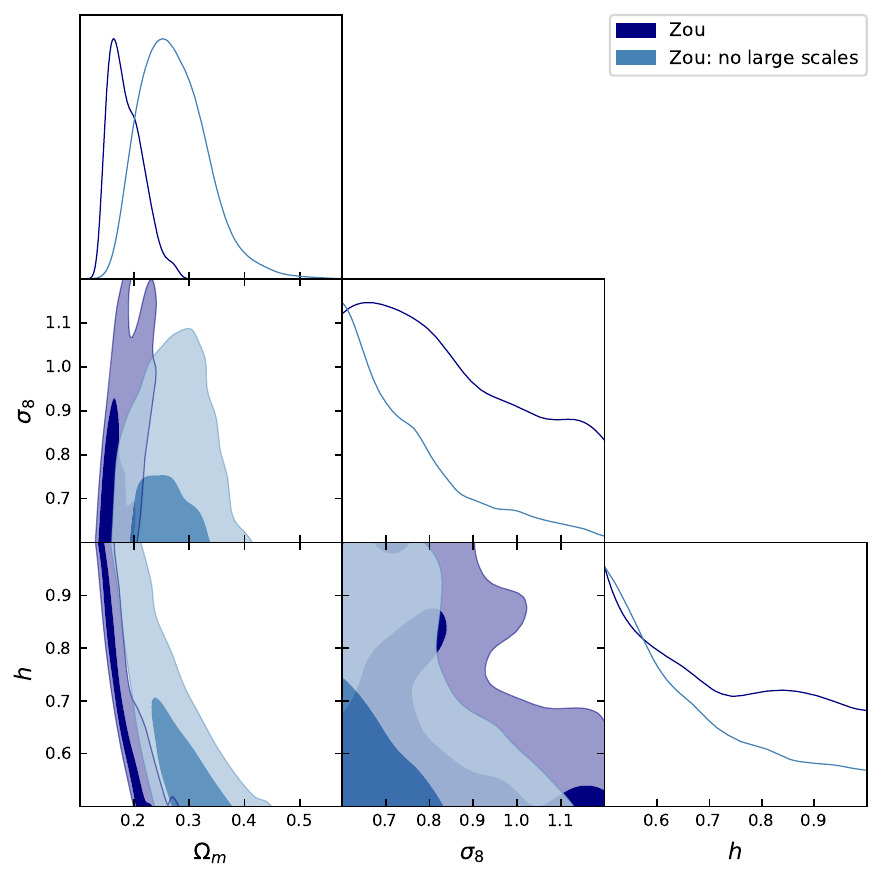}
     \caption{Marginalised posterior distributions and probability contours for the cosmological parameters for the full data case (dark blue) and excluding the large-scale data (light blue).
     }
     \label{fig:corner_zou}
\end{figure}

\begin{table*}[ht]
  \caption{
  Best-fit parameters and  68\% confidence interval determined using the ZOU cluster catalogue. 
  }
  \label{tab:zou_results}
  \centering
  \begin{tabular}{ccccccc} 
    \hline\hline
    \multicolumn{1}{c}{} & \multicolumn{3}{c}{ZOU} & \multicolumn{3}{c}{ZOU: no large scales} \\
    \cmidrule(r){2-4} \cmidrule(r){5-7}
    Parameter & Mean & Mode & 68\% CI & Mean & Mode & 68\% CI\\
    \midrule
    $\log{M_{min}}$ & 
       13.18 & 13.15 & [13.09 , 13.28] &
       13.40 & 13.43 & [13.29 , 13.53] \\
    $\log{M_{1}}$ &
       14.03 & 13.55 & [13.25 , 14.33] &
       14.07 & 13.79 & [13.48 , 14.27] \\
    $\alpha$ &
       0.76 & - & [0.50 , 0.81] &
       0.75 & - & [0.50 , 0.80] \\
    $\Omega_m$ & 
       0.19 & 0.16 & [0.15 , 0.21] &
       0.27 & 0.25 & [0.20 , 0.32] \\
    $\sigma_8$ &
       0.85 & - & [0.60 , 0.94] &
       0.76 & - & [0.60 , 0.79] \\
    $h$ &
       0.72 & - & [0.50 , 0.81] &
       0.67 & - & [0.50 , 0.71] \\
    $\beta$ & 
       2.79 & 2.80 & [2.69 , 2.89] &
       2.79 & 2.70 & [2.79 , 2.90] \\
    \hline\hline
  \end{tabular}
  \tablefoot{On the left side, results obtained from the entire dataset are presented. On the right side, results obtained after excluding the large-scale cross-correlation data points are shown.}
\end{table*}


Figure \ref{fig:sampled_zou}  illustrates the posterior sampling of the cross-correlation function for the ZOU dataset. Overall, both scenarios exhibit a remarkable consistency between the best fit and the sampled region when compared to the data. However, noticeable differences emerge between them at the largest scales. 

Figure \ref{fig:zou_hod} represents the marginalised posterior distribution of the HOD parameters for both datasets. The parameters $\log M_{\text{min}}$ and $\log M_1$ exhibit relatively robust constraints with mode values\footnote{Throughout this study, all masses are expressed in units of $M_{\odot}/h$.} of 13.15 and 13.55, respectively. As it will be discussed in more detail later, these values are reasonable considering the mean mass and richness for the galaxy clusters from the ZOU catalogue.
The inclusion of large-scale data significantly impacts $\log M_{\text{min}}$, leading to a shift in the posterior sampling towards slightly lower values. Specifically, the mode value decreases from 13.43 to 13.15 when incorporating the large-scale data. In the case of $\log M_1$ the changes are not relevant from a statistical point of view.
Conversely, for the parameter $\alpha$, we were only able to establish an upper limit\footnote{All upper limits are provided at a 68\% confidence level.}. In both scenarios, with and without the inclusion of large-scale data, the upper limit of $\alpha$ remain consistent at 0.81 and 0.80, respectively.

Figure \ref{fig:corner_zou} displays the marginalised posterior distributions and probability contours\footnote{The probability contours in the corner plots are set to 0.393 and 0.865 by default.} for the estimated cosmological parameters. For a more comprehensive view, please refer to the complete corner plot in Appendix \ref{app:corner plots} (Fig. \ref{fig:Zou_full_corner}). Our data do not provide tight constraints for either the parameter $h$ or $\sigma_8$, resulting in upper limits for both. Specifically, the upper limits for $h$ are 0.81 and 0.71 for the complete dataset and the dataset excluding large scales, respectively. As for $\sigma_8$, the upper limits are 0.94 and 0.79 in the same scenarios. These results are generally well aligned with the most recent and precise measurements\footnote{In this study, the \textit{Planck} 2018 results, presented in \cite{PLA18_VI}, are used as reference points. Namely, $\Omega_m=0.315\pm0.007$, $\sigma_8=0.811\pm0.006$ and $H_0 = (67.4\pm0.5)\,\text{km}\,\text{s}^{-1}\,\text{Mpc}^{-1}$.}. 
However, in contrast, $\Omega_m$ is well constrained by both datasets. Yet, the anomaly observed at large scales has a notable impact on its determination, resulting in a significant increase in its value (from a mode of 0.16 to 0.25), corroborating similar findings by \citet{Cue23}. When we exclude the anomalous large-scale data points, the predictions yield broader ranges of values for $\Omega_m$ ($0.19^{+0.02}_{-0.04}$ and $0.27^{+0.05}_{-0.07}$, respectively), but with a peak that aligns more closely with the \textit{Planck} value of $\Omega_m = 0.315$.


\begin{figure*}[ht]
    \centering
    \begin{subfigure}{0.5\textwidth}
        \includegraphics[width=\linewidth]{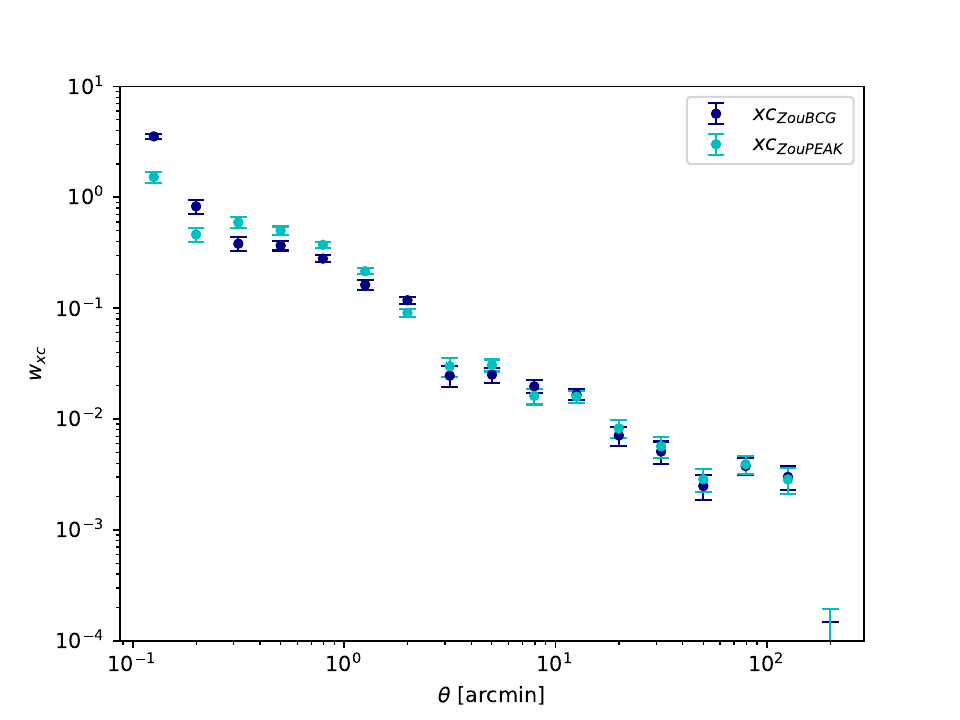}
        \caption{}
    \end{subfigure}%
    \begin{subfigure}{0.5\textwidth}
        \includegraphics[width=\linewidth]{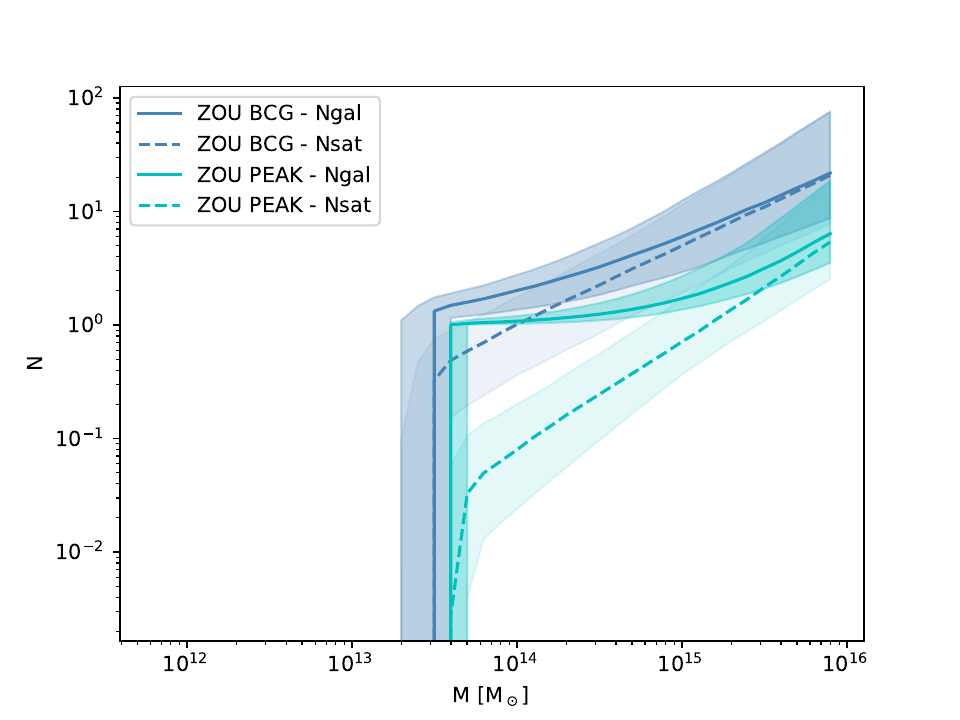}
        \caption{}
    \end{subfigure}
    \caption{Left: Cross-correlation data obtained for the ZOU cluster with BCG positions (dark blue) and PEAK positions (light blue). Right: Sampling plot illustrating the relationship between the total number of galaxies ($N_{cen}+N_{sat}$, solid line) or the number of satellites ($N_{sat}$, dashed line) and the halo mass (M) based on the HOD model for the ZOU cluster. Dark blue represents data obtained with BCG positions, while light blue represents data obtained with PEAK positions. The solid and dashed lines depict the sampling median for the number of galaxies and satellites, respectively, and the shaded regions indicate the 68\% confidence interval. Large-scale data were excluded in this particular analysis.
    } 
    \label{fig:HOD_peakbcg}
\end{figure*}

As explained in Sect. \ref{subsec:data}, two distinct central positions are available for the clusters in the ZOU catalogue. Consequently, an additional study was carried out to assess the potential impact of this choice on the results. Fig. \ref{fig:HOD_peakbcg} displays both the cross-correlation data (left) and a sampling of the HOD obtained using the BCG (dark blue) and PEAK (light blue) central positions. For the analysis, large-scale data were discarded for both datasets. 

The cross-correlation function reveals a strong concordance between the two datasets for angular separations exceeding $\sim 3$ arcmin, as expected. However, divergences emerge at smaller scales, where the HOD is most significantly affected. Evidence supporting this assertion is presented in the right panel of Fig. \ref{fig:HOD_peakbcg}. This plot shows the overall number of galaxies (solid line) and the count of satellite galaxies (dashed line) in relation to the host halo mass, according to the adopted HOD model. The solid and dashed lines represent the medians derived from the sampling of total galaxies and satellites, respectively, with shaded regions indicating the 68\% confidence intervals. 

The satellite count is significantly higher in the BCG dataset, reaching a maximum of around 100 (with a 68\% confidence interval) and a median slightly exceeding 20. In contrast, the PEAK dataset shows a maximum of 10 satellites (with a median of 3). This difference is interesting and illustrates how the HOD can be exploited to evaluate the feasibility of the obtained fit. The ZOU catalogue has a median richness of 22.5, with a lower limit of 10 imposed by the cluster detection algorithm \citep{ZOU21}. Consequently, results obtained from the PEAK-position dataset can be dismissed, as they inadequately reflect the actual physical observations. Conversely, the BCG dataset provides a number of satellites that are in better agreement with the expected outcome. 

A comprehensive corner plot, providing a comparison of the posterior probability distribution for both datasets, can be found in Fig. \ref{fig:zou_peak_corner} within Appendix \ref{app:corner plots}. Significant variations become evident across several parameters, with the most significant differences observed within the HOD, as previously discussed. Concerning the cosmological parameters, the main difference is in $\sigma_8$, which must be adjusted to compensate for the considerably higher value of $\log M_1$ in the PEAK case. It is worth noting that a higher $\log M_1$ value implies a lower number of satellite galaxies.

\subsection{WEN catalogue}
\label{subsec:wen}

This subsection presents the results derived from the WEN catalogue analysis. Once again, two different scenarios are explored by considering or not the largest scale data. While in the case of the ZOU catalogue, data points above $\gtrsim 60$ arcmin were excluded, for the WEN catalogue, an additional point was removed due to being considered an outlier, as discussed above. For further information, one should refer to Figs. \ref{fig:xc_data} and \ref{fig:xc_mpc} and the corresponding discussion.

\begin{figure}[ht]
\includegraphics[width=0.5\textwidth]{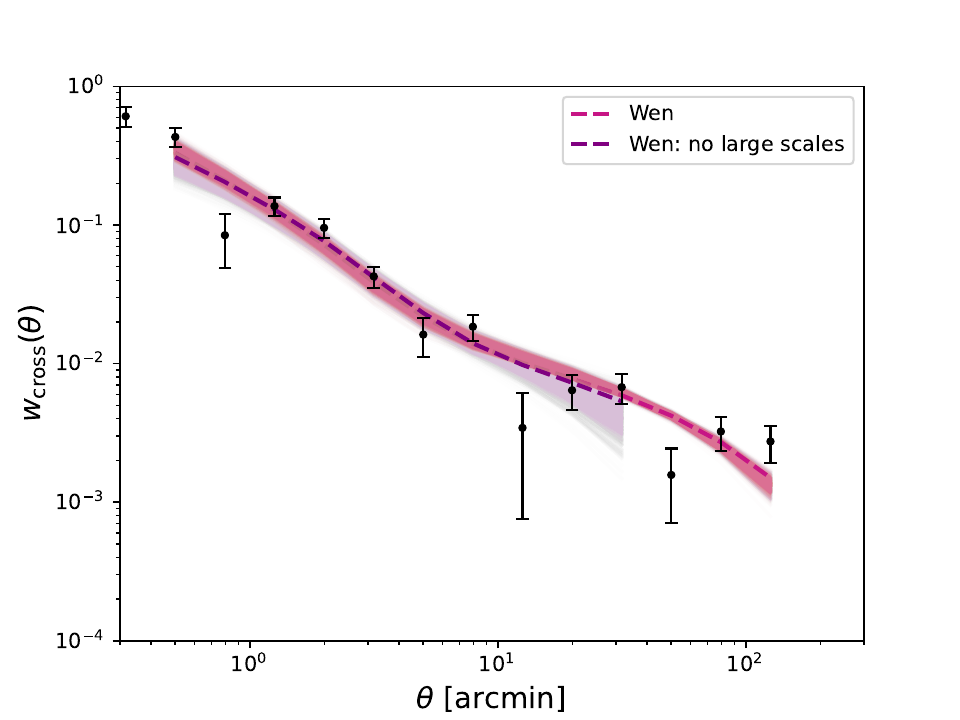}
 \caption{Sampling plots for WEN using all points and excluding large-scale data. The pink and purple regions denote all data points and the excluded large-scale measurements, respectively. The dashed lines depict the best-fit results, and the data are represented by black dots.
 }
 \label{fig:sampled_wen}
\end{figure}

\begin{figure*}[ht]
    \centering
    \includegraphics[width=0.32\textwidth]{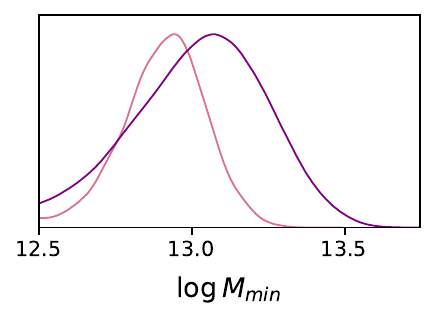}
    \includegraphics[width=0.32\textwidth]{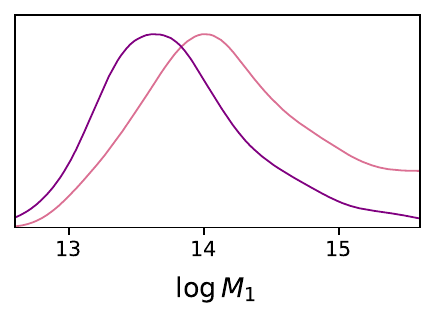}
    \includegraphics[width=0.32\textwidth]{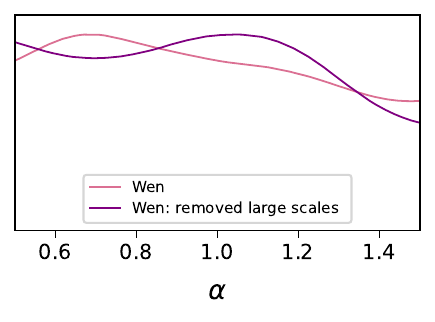}
    \caption{Marginalised posterior distribution for the HOD. The measurements from the complete dataset are represented in pink, while the measurements with removed large-scale data are presented in purple.}
    \label{fig:wen_hod}
\end{figure*}

\begin{figure}[htbp]
\includegraphics[width=0.5\textwidth]{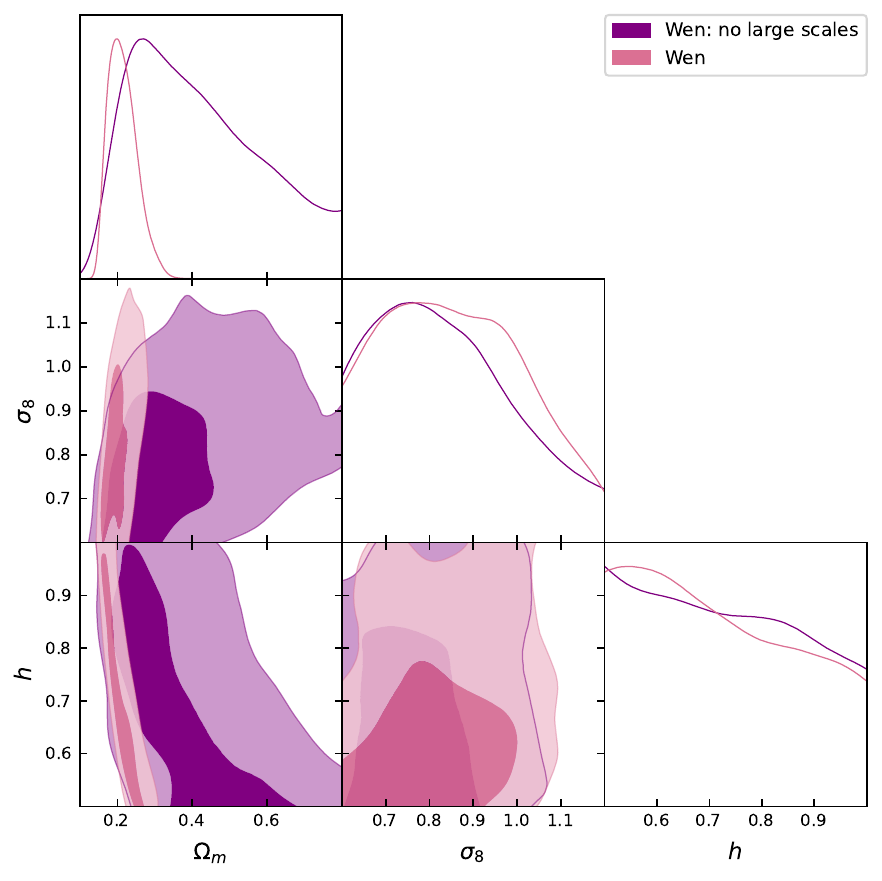}
 \caption{Marginalised posterior distributions and probability contours for the cosmological parameters for the full data (pink) and the removed large-scale data (purple).
 }
 \label{fig:cosmo_wen}
\end{figure}

\begin{table*}[ht]
  \caption{
  Best-fit parameters and  68\% confidence interval determined using the WEN cluster catalogue. }
  \label{tab:wen_results}
  \centering
  \begin{tabular}{ccccccc} 
    \hline\hline
    \multicolumn{1}{c}{} & \multicolumn{3}{c}{WEN cluster: all data} & \multicolumn{3}{c}{WEN cluster: no large scales} \\
    \cmidrule(r){2-4} \cmidrule(r){5-7}
    Parameter & Mean & Mode & 68\% CI & Mean & Mode & 68\% CI\\
    \midrule
    $\log{M_{min}}$ & 
       12.91 & 12.94 & [12.75 , 13.09] &
       13.03 & 13.07 & [12.83 , 13.27] \\
    $\log{M_{1}}$ &
       14.19 & 14.00 & [13.43 , 14.73] &
       13.81 & 13.62 & [13.14 , 14.20] \\
    $\alpha$ &
       0.97 & - & [0.50 , 1.50] &
       0.97 & - & [0.5 , 1.13] \\
    $\Omega_m$ & 
       0.21 & 0.20 & [0.17 , 0.24] &
       0.41 & 0.27 & [0.18 , 0.52] \\
    $\sigma_8$ &
       0.86 & 0.77 & [0.66 , 0.99] &
       0.84 & 0.76 & [0.62 , 0.93] \\
    $h$ &
       0.72 & - & [0.50 , 0.80] &
       0.73 & - & [0.50 , 0.81] \\
    $\beta$ & 
       2.80 & 2.81 & [2.71 , 2.90] &
       2.79 & 2.81 & [2.70 , 2.90] \\
    \hline\hline
  \end{tabular}
  \tablefoot{On the left side, results obtained from the entire dataset are presented. On the right side, results obtained after excluding the large-scale cross-correlation data points are shown.}
\end{table*}

Table \ref{tab:wen_results} summarises the obtained parameter values in both scenarios. The complete corner plot can be found in Appendix \ref{app:corner plots}, Figure \ref{fig:wen_corner}. Figure \ref{fig:sampled_wen} illustrates the best-fit results and the posterior sampling of the cross-correlation function, both with and without the large-scale data. As observed in the ZOU catalogue analysis, the best-fit results and sampled regions align well with the observed data.

Figure \ref{fig:wen_hod} displays the marginalised posterior distribution of the HOD parameters for both datasets. The results are consistent in both cases. The parameters $\log M_{\text{min}}$ and $\log M_1$ exhibit peaked posterior distributions with values that align reasonably well with the mean mass and richness from the WEN catalogue. The broader distributions compared to the ZOU results may be attributed to the relatively low amount of statistics available. The removal of large-scale data leads to slight shifts in both $\log M_{\text{min}}$ and $\log M_1$ parameters. Interestingly, while the mode value of $\log M_{\text{min}}$ increases from 12.94 to 13.07, the mode of $\log M_1$ decreases from 14.00 to 13.62, in contrast to the ZOU catalogue analysis. The $\alpha$ parameter remains unconstrained, and no definitive conclusions can be drawn.

Figure \ref{fig:cosmo_wen} presents the marginalised posterior distribution and probability contours for all estimated cosmological parameters. Once again, the posterior distributions are broader compared to the ZOU results. When the large-scale data are excluded, only $\Omega_m$ is notably affected, shifting towards higher values more in line with traditional expectations (the mode shifts from 0.20 to 0.27), similar to the ZOU analysis. In both cases, $\sigma_8$ exhibits a broad peaked distribution with a mode around 0.8. An upper limit (68\% CI) of $\sim0.8$ was obtained for $h$ in both cases.

\subsection{Discussion}

In this section, the findings obtained from the ZOU and WEN catalogues are compared with the latest results derived from employing a galaxy catalogue as lenses for studying magnification bias. Additionally, we compared our findings with the latest constraints obtained by weak-lensing studies such as the Kilo-Degree Survey \citep[KiDS;][]{KUI19} and the Dark Energy Survey \citep[DES;][]{DES16}. Despite the definition differences between the two cluster catalogues, particularly in terms of the number of sources and average mass and richness, the parameter constraints derived from both catalogues exhibit good mutual agreement and alignment with the current consensus values (see Tables \ref{tab:zou_results} and \ref{tab:wen_results}).

\begin{figure}[ht]
\includegraphics[width=0.5\textwidth]{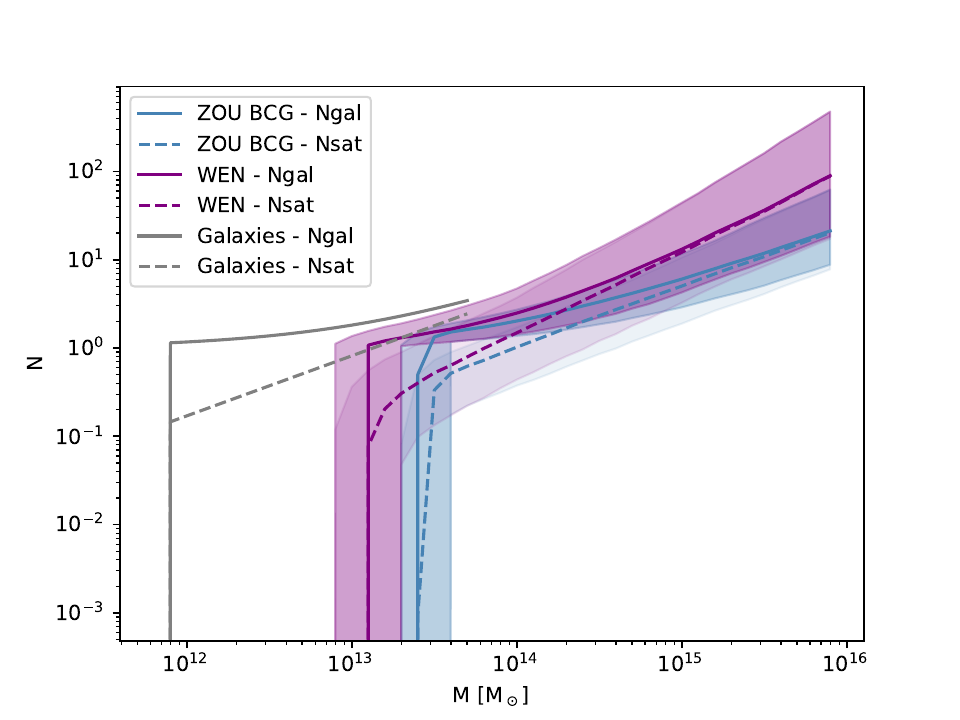}
 \caption{Mapping the total number of galaxies ($N_{cen}+N_{sat}$, solid line) or the number of satellites ($N_{sat}$, dashed line) to halo mass (M) using the HOD model across various lenses in this study. Samplings for the clusters are depicted in blue (ZOU) and purple (WEN); the solid and dashed lines represent the sampling median for the number of galaxies and satellites, respectively, and the shaded regions indicate the 68\% confidence interval. Large-scale data were excluded for both datasets. Additionally, results obtained by applying the HOD model to mean data from the galaxy sample in \cite{Cue23} are shown in grey.  
 }
 \label{fig:HOD_allcases}
\end{figure}

In the context of HOD results, Fig. \ref{fig:HOD_allcases} provides a summary of the HOD outcomes obtained for both the ZOU and WEN catalogues, alongside galaxies from \cite{Cue23} (grey). Notably, the estimated minimum halo mass, $\log M_{\text{min}}$, exceeds $\gtrsim 13$ for both cluster catalogues. These estimates are consistent with expectations for typical large red galaxies (LRGs), which are akin to BCGs. The modelling of the LRG angular correlation function suggests an effective halo mass of $M_{\text{eff}}=2.8-4.4\times10^{13}M_{\odot}$ \citep{BLA08}, and compatible masses are also derived from the analysis of their large-scale redshift-space distortions \citep[$M=3.5^{+1.8}_{-1.4}\times10^{13}M_{\odot}$][]{CAB09, BAU14}. Additionally, these estimates align with central masses derived for galaxy clusters using an independent methodology based on magnification bias at small angular separations, enhanced by stacking techniques \citep[see also][]{FER22, Cre22, CRE24}.

Furthermore, it is noteworthy that the minimum mass estimated for the WEN catalogue is smaller than that for the ZOU catalogue, consistent with findings in \citet{CRE24}. This difference may stem from variations in the average total mass of both catalogues, with $M=1.23\times10^{14}M_{\odot}$ for ZOU and $1.00\times10^{14}M_{\odot}$ for WEN. In both cases, these masses are approximately an order of magnitude higher than those obtained for the galaxy sample. This discrepancy underscores that our usage of galaxy clusters as gravitational lenses introduces an obvious selection towards higher-mass structures.

In terms of the number of galaxy satellites, one would naturally expect this number to be significantly higher in the case of galaxy clusters compared to individual galaxies. Previous studies in this context have consistently indicated that galaxies serving as gravitational lenses tend to be either isolated or the most massive galaxies within small groups \citep[e.g.][]{BON19, BON20, Cue23}, as corroborated by the shape of the HOD curve. 

However, it is essential to note that, even accounting for uncertainties in the estimated parameters, they predict a relatively low number of satellites for galaxy clusters, considering the mean richness of both cluster catalogues. This result suggests that a significant portion of the magnification bias lensing effect is primarily driven by the BCGs themselves, possibly with some contribution from secondary massive satellite galaxies. Consequently, the collective halo mass of the entire galaxy cluster appears to play a less substantial role than the individual mass of the BCGs and a few massive satellites. This observation may be related to the considerably higher mass concentration in the inner regions of the cluster, as opposed to the outer regions \citep[see, for example,][]{Cre22}.

Moreover, despite the WEN catalogue having a lower average richness, it yields a higher estimated number of satellites compared to the ZOU catalogue. This discrepancy raises the possibility of an estimation issue with the $M_1$ or $\alpha$  parameter, or both, as indeed is the case. This observation underscores the potential utility of the HOD representation as an additional tool for result validation.

Turning our attention to the analysis of cosmological parameters, a key observation, as previously emphasised, is the influence of the large-scale excess in cross-correlation on cosmological constraints. This anomaly is evident in both cluster datasets and leads to a notable shift in the behaviour of the $\Omega_m$ parameter. While it exhibits a peak around 0.30, this peak recedes to approximately 0.20 when incorporating the large-scale data into all the analysed datasets.

\begin{figure}[htbp]
\includegraphics[width=0.5\textwidth]{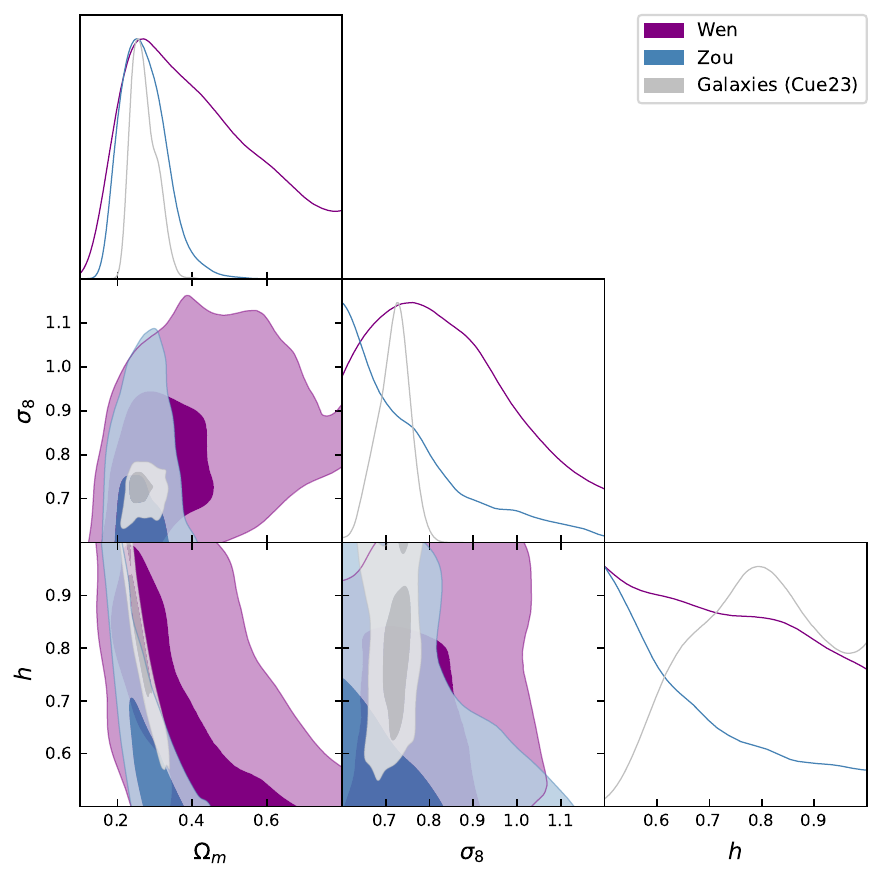}
 \caption{Marginalised posterior distributions and probability contours for cosmological parameters for the ZOU cluster (in blue) and WEN cluster (in purple). Results using the galaxy sample from \cite[specifically, the case excluding the GAMA 15 region]{Cue23} are depicted in grey. Large-scale data were excluded in the analysis for both clusters. 
 }
 \label{fig:cosmo_all}
\end{figure}

Figure \ref{fig:cosmo_all} illustrates the marginalised posterior distribution and probability contours for the cosmological parameters obtained from the ZOU and WEN catalogues, in comparison to the galaxy analysis conducted by \cite{Cue23}. The analysis excludes the large-scale data from the cross-correlation function for the cluster samples, and the GAMA 15 region for the galaxies. As previously stated, the constraints established on cosmological parameters remain consistent across both cluster datasets. While the WEN catalogue yields broader parameter ranges, it effectively constrains $\sigma_8$, in contrast to the upper limit provided by the ZOU catalogue.

These findings are consistent with those reported by \cite{Cue23}, who employed galaxies as gravitational lenses. The larger dataset of galaxies and peaked redshift distribution contribute to more precise constraints on all the examined cosmological parameters. Notably, the galaxy sample yields a tentative peak in the $h$ parameter, whereas the cluster-based results only provide rough upper limits.

\begin{figure}[htbp]
\includegraphics[width=0.5\textwidth]{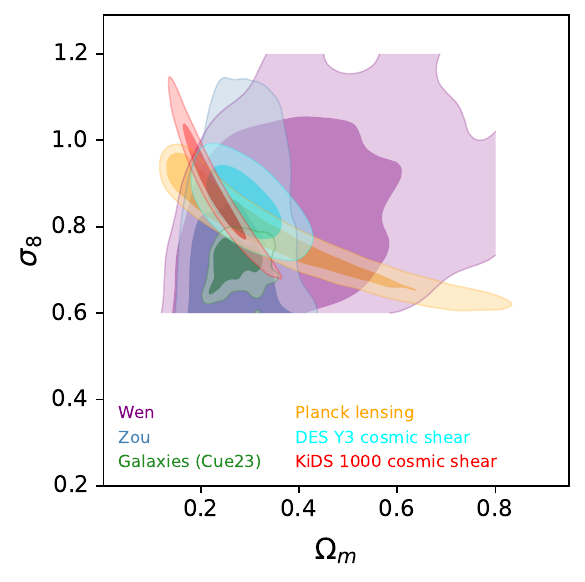}
 \caption{Comparison of contour plots in the $\Omega_m - \sigma_8$ parameter space sourced from various studies. The outcomes of our study are represented in purple (WEN cluster) and dark blue (ZOU cluster). In addition, we show the outcomes of previous investigations within our group using the same methodology on a galaxy sample denoted as CUE23 and depicted in green. External datasets for comparison include \textit{Planck} lensing CMB results in yellow, DES Year 3 cosmic shear results in light blue, and KiDS 1000 cosmic shear results in red. Notably, in contrast to the rest of this paper, the contour levels here are set at 0.68 and 0.95.
 }
 \label{fig:omgMsgm8}
\end{figure}

The results from our study agree with the findings of prominent large-scale collaborations that employ weak lensing as a tool for probing cosmology, as shown in Fig. \ref{fig:omgMsgm8}. Specifically, we compare our results with those from the European Southern Observatory KiDS\footnote{http://kids.strw.leidenuniv.nl (chains: KiDS 1000 xipm)}, the DES\footnote{https://www.darkenergysurvey.org (chains: 1x2pt lcdm SR maglim)}, and the \textit{Planck} collaboration\footnote{https://www.cosmos.esa.int/web/planck/pla (chains: base lensing lenspriors)}. While these collaborations use a combination of different observables in their cosmological measurements, for this comparison, we specifically focused on the aspects most closely related to our study. In the case of DES Year 3 data release, we considered cosmic shear measurements \citep{DES22}. Similarly, for the KiDS-1000 data release, we concentrated on cosmic shear data \citep{KIDS21}. Additionally, we incorporated \textit{Planck}'s CMB lensing results from the 2018 data release \citep{PLA18VIII}.

In Fig.\ref{fig:omgMsgm8}, contour plots of the parameter space $\Omega_m - \sigma_8$ for various studies are superimposed. It can be observed that the cluster results, while producing broad distributions (especially in the case of WEN), are in agreement with the results from \textit{Planck}, DES, and KiDS. This demonstrates that the study of the magnification bias produced by clusters, while currently not statistically competitive due to the limited samples used in the present work, provides results that reinforce the values for cosmological parameters obtained in other studies.

It is worth noting that, as with other studies testing the local Universe, the results appear to trend towards values of $\sigma_8 < 0.8$ and $\Omega_m< 0.3$ in slight `tension' with the more precise value from \textit{Planck} CMB. However, it is essential to significantly improve the precision of this analysis and evaluate its possible systematic errors, such as identifying the origin of the excess cross-correlation signal at high angular scales, before using it to test the validity of the $\Lambda$ cold dark matter theory. In fact, the latest results from DES (Year 3 data release) reduce this tension compared to the Year 1 data release and find values compatible with \textit{Planck} CMB \citep{DES22}. More statistical data and a deeper understanding of potential data analysis errors, as well as alternative tests of the universe at high redshifts, are necessary to advance in this direction.

These results are promising for the utilisation of clusters as gravitational lenses within this methodology. Remarkably, despite the significantly smaller number of sources compared to the galaxy sample, comparable results were achieved. This further underscores the robustness of the methodology, which demonstrates consistency across different lens samples.

\section{Conclusions}
\label{sec:concl}

This study examines, for the first time, the magnification bias effect exerted on SMGs by galaxy clusters. It uses two different galaxy cluster samples as lenses: the ZOU catalogue, with about 9000 clusters, and the WEN catalogue, with over 3600 clusters. The background sample is derived from the H-ATLAS high-redshift SMGs within the GAMA fields (G09, G12, and G15) and the NGP, and it comprises almost 5700 sources.
The primary focus was on analysing the cross-correlation function to determine constraints on cosmological and astrophysical parameters through the use of a MCMC algorithm. 

The cross-correlation function and covariance matrix were calculated following the methodology detailed in \cite{GON23} and summarised in \cite{Cue23}. This was done using a modified version of the estimator introduced by \cite{LAN93} and adapted by \cite{HER01}. The results for both clusters are higher than those for galaxies. Furthermore, we identified an excess at larger scales, similar to what is seen with galaxies, hinting at a potential underlying physical cause (such as contributions from additional large-scale structures). 

When examining the relationship between cross-correlation and physical distance, the two cluster datasets, even though they originate from unrelated catalogues, exhibit a similar pattern. This similarity is particularly noticeable in the transition region from the one-halo to two-halo regime, as well as the presence of an anomalous bump at larger scales. At smaller scales within the one-halo regime, there is a slight difference in normalisation between the data from the two cluster samples, which could be linked to the richness of each catalogue. 

In the ZOU sample, two different cluster positions were provided: one based on the BCG sample and the other on the PEAK sample. After careful investigation, we have reached the conclusion that using the PEAK option does not yield realistic results for the HOD analysis.

The results obtained from the two cluster samples are consistent and in agreement with the findings reported by \cite{Cue23} from the galaxy sample. However, it is important to note that the parameters derived from the cluster samples are less constrained compared to the galaxy sample. This is attributed to the wider redshift distributions and the lower number of clusters available for analysis in comparison to the galaxies, which limits the precision of our parameter estimation.

Moreover, our study's results align with the latest results from well-established large-scale collaborations that employ weak lensing for cosmological exploration. As with other low-$z$ cosmological studies, we find that our results (although they exhibit broad distributions due to limited statistics) appear to point towards lower values of $\sigma_8 < 0.8$ and $\Omega_m < 0.3$, in slight tension with the more precise data from \textit{Planck}. While they do not yield competitive constraints to endorse or reject the $\Lambda$ cold dark matter model, this result underscores the necessity for improved precision, systematic error assessment, and alternative tests of the high-redshift universe.

In conclusion, despite the relatively low amount of available lenses, reasonable constraints, consistent with prevailing consensus values, were successfully obtained by studying the magnification bias produced by galaxy clusters on SMGs. This achievement was made possible by capitalising on the elevated mass (and thus impact) of entire clusters in contrast to individual galaxies in the lensing process. This highlights the potential of utilising galaxy clusters as lenses in magnification bias studies, presenting a promising and independent avenue for constraining cosmological parameters. Further advancement in this approach can be made by expanding the number of lenses under study, for example by increasing the overlapping area of background sources with the cluster catalogues.

\begin{acknowledgements}
RFF, LB, DC, JGN and JMC and acknowledge the PID2021-125630NB-I00 project funded by MCIN/AEI/10.13039/501100011033/FEDER, UE.
LB also acknowledges the CNS2022-135748 project funded by MCIN/AEI/10.13039/501100011033 and by the EU “NextGenerationEU/PRTR”.
JMC also acknowledges financial support from the SV-PA-21-AYUD/2021/51301 project.\\
We deeply acknowledge the CINECA award under the ISCRA initiative, for the availability of high performance computing resources and support. In particular the projects `SIS22\_lapi', `SIS23\_lapi' in the framework `Convenzione triennale SISSA-CINECA'.\\
The \textit{Herschel}-ATLAS is a project with \textit{Herschel}, which is an ESA space observatory with science instruments provided by European-led Principal Investigator consortia and with important participation from NASA. The H-ATLAS web- site is http://www.h-atlas.org. GAMA is a joint European- Australasian project based around a spectroscopic campaign using the Anglo- Australian Telescope. The GAMA input catalogue is based on data taken from the Sloan Digital Sky Survey and the UKIRT Infrared Deep Sky Survey. Complementary imaging of the GAMA regions is being obtained by a number of independent survey programs including GALEX MIS, VST KIDS, VISTA VIKING, WISE, \textit{Herschel}-ATLAS, GMRT and ASKAP providing UV to radio coverage. GAMA is funded by the STFC (UK), the ARC (Australia), the AAO, and the participating institutions. The GAMA web- site is: http://www.gama-survey.org/.\\
This research has made use of the python packages \texttt{ipython} \citep{ipython}, \texttt{matplotlib} \citep{matplotlib} and \texttt{Scipy} \citep{scipy}.
\end{acknowledgements}

\bibliographystyle{aa} 
\bibliography{xc_clusters} 

\begin{appendix}

\section{Theoretical framework}
\label{app:framework}

 The SMG magnification bias offers a novel and promising approach to constraining cosmology \citep[][]{GON17, BON20, CUE21, GON21}. This method leverages a weak-lensing-induced cross-correlation, employing a sample of foreground galaxies and a background set of SMGs. Magnification bias, a well-documented phenomenon \citep[see][and references therein]{Bar01}, leads to an enhancement in the flux and solid angle of distant sources. However, when a flux threshold is applied, a mismatch arises between these effects, resulting in an excess of background sources around the foreground ones, referred to as lenses, compared to situations without lensing influence.

This work focuses on studying the weak-lensing-induced foreground-background number cross-correlation as its main observable. The correlation arises from the impact of weak lensing caused by the mass density field associated with foreground galaxy clusters on the number counts of the background galaxy sample through magnification bias explained in \citep{Cue23}. To compute this correlation, the study employs the halo model formalism proposed by \cite{COO02}, along with the Limber and flat-sky approximations. The correlation can be evaluated with the following equation: 

\begin{equation}
    \label{eq:w_fb}
    w_{\text{fb}}(\theta)=2(\beta -1)\int^{\infty}_0 \frac{dz}{\chi^2(z)}\frac{dN_f}{dz}W^{lens}(z)\int_{0}^{\infty}\frac{ldl}{2\pi}P_{\text{g-m}}(l/\chi^2(z),z)J_0(l\theta) 
,\end{equation}where $W^{\text{lens}}(z)$ is defined as
\begin{equation}
W^{\text{lens}}(z)=\frac{3}{2}\frac{H_0^2}{c^2}E^2(z)\int_z^{\infty} dz' \frac{\chi(z)\chi(z'-z)}{\chi(z')}\frac{dN_b}{dz'}
.\end{equation}Here, $E(z)=\sqrt{\Omega_m(1+z)^3+\Omega_{\Lambda}}$, $dN_b/dz$ and $dN_f/dz$ are the unit-normalised background and foreground redshift distributions, $\chi(z)$ is the comoving distance to redshift z and $P_{\text{g-m}}$ is the galaxy-matter cross-power spectrum. $\beta$ represents the logarithmic slope of the background sources' number counts, and a detailed discussion of this parameter can be found in \citet{Cue23}. 

To reduce the computational complexity in evaluating the model due to the substantial number of integrals required for each computation, we developed a mean-redshift approximation. This technique avoids the direct calculation of the outermost integrals in Eqs. \ref{eq:w_fb} over the entire redshift distribution. Instead, it focuses on the sample's mean redshift. It has been validated for galaxies in \citet{Cue23} and will also be examined for clusters in the results' section. The corresponding expression for the foreground-background cross-correlation under the redshift approximation is

\begin{equation}
\label{eq:w_fb_approx}
    w_{\text{fb}}(\theta)\approx 2(\beta-1)\frac{W^{\text{lens}}(\bar{z})}{\chi^2(\bar{z})}\int_0^{\infty}\frac{ldl}{2\pi}P_{\text{g-m}}(l/\chi(\bar{z}),\bar{z})J_0(l\theta).
\end{equation}

As described in \citet{Cue23}, the cross-power spectrum of galaxies and matter is used in cosmology to study the relationship between the spatial distribution of galaxies and the underlying matter distribution in the universe. It stems from the previously introduced halo model formalism, where all mass in the universe is grouped into distinct units called halos. The clustering of matter within halos dominates correlations on small scales, whereas only the spatial distribution of halos becomes important on larger scales. This results in the cross-power spectrum being composed of two contributions: the one-halo term and the two-halo term. The one-halo term primarily governs galaxy-matter correlations within individual halos, dominating at small angular distances, and it quantifies the clustering of galaxies within the same dark matter halo. On the other hand, the two-halo term accounts for cross-correlations between different dark matter halos, becoming more relevant on larger scales, typically beyond the scale of an individual halo (around 1 Mpc). 

The dependence of the cross-correlation function on the value of beta and the cosmological parameters of the assumed underlying model is influenced by the way galaxies populate dark matter halos. This population is described by the HOD. In this work, the simple three-parameter model introduced by \cite{ZHE05} is employed for this purpose. According to this model, a galaxy is located at the centre of a halo when its mass exceeds a certain threshold, $M_{min}$. Any additional galaxies are treated as satellites, and their distribution directly follows the halo's mass profile, as detailed in works such as \cite{ZHE05}. When a halo's mass exceeds a different threshold, $M_1$, it will host satellites, and the number of satellites present is described by a power-law function with a coefficient of $\alpha$. Consequently, the probability of a central galaxy being present is represented as a step function:

\begin{equation}
    \label{eq:ncen01}
    N_\text{c}(M) =
    \begin{cases}
    0 \quad \text{if}\ M < M_\text{min}\\
    1 \quad \text{otherwise}
    \end{cases}
.\end{equation}
The satellite galaxies occupation can be described as
\begin{equation}
    \label{eq:nsat01}
    N_\text{s}(M) = N_\text{c}(M) \cdot \biggl(\dfrac{M}{M_1}\biggr)^{\alpha}
,\end{equation}where $M_\text{min}$, $M_1$, and $\alpha$ are the free-parameters of the model.

\clearpage
\onecolumn
\section{Corner plots}

\label{app:corner plots}
\begin{figure}[h]
  \centering
  \includegraphics[width=1\textwidth]{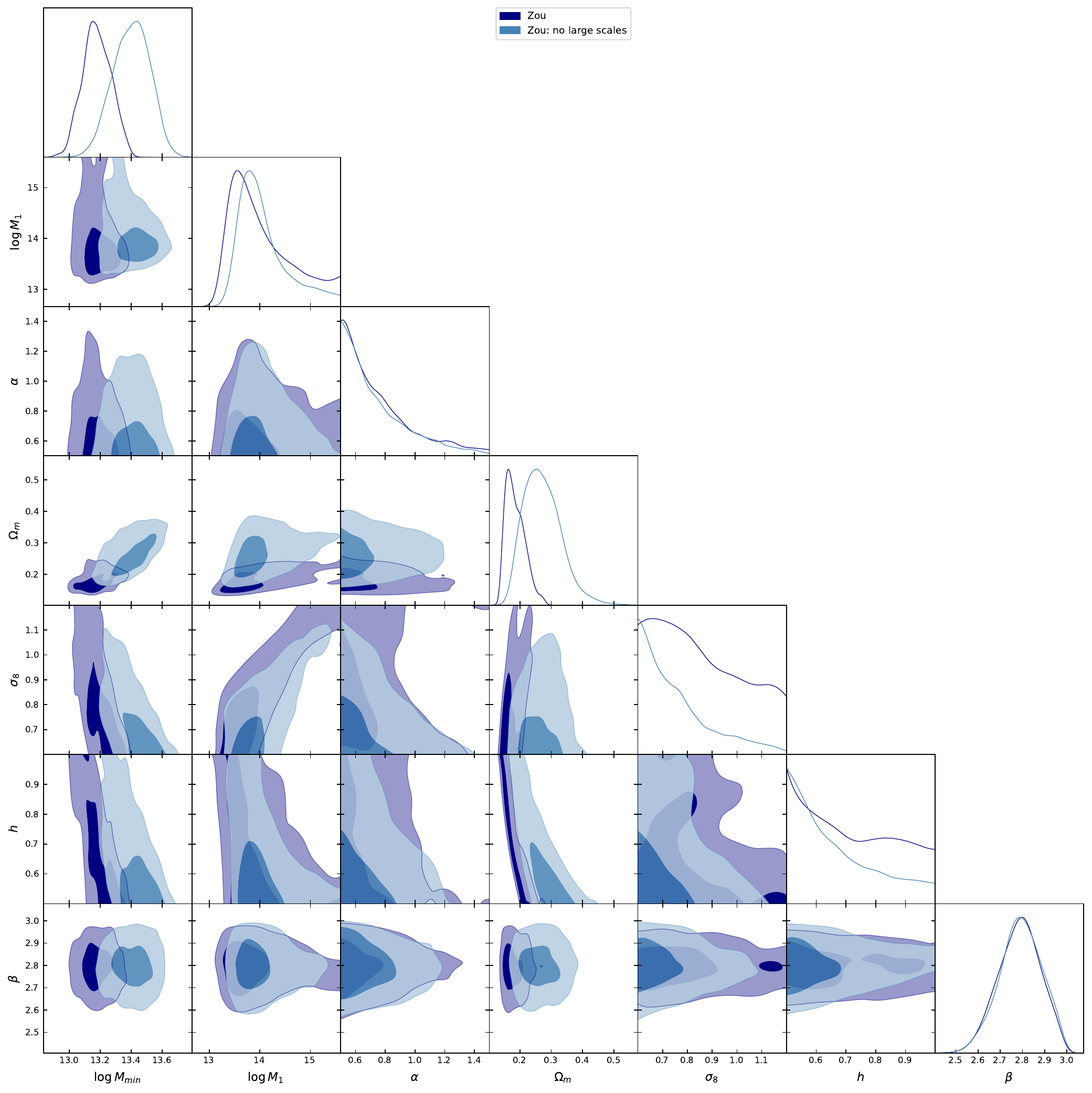}
  \caption{Marginalised posterior distributions and probability contours for the MCMC runs on the cross-correlation function using ZOU cluster data. The dark blue fit was obtained using all available data. The light blue fit was obtained excluding cross-correlation data above $\gtrsim 60$ arcmin.}
  \label{fig:Zou_full_corner}
\end{figure}

\begin{figure}[h]
  \centering
  \includegraphics[width=1\textwidth]{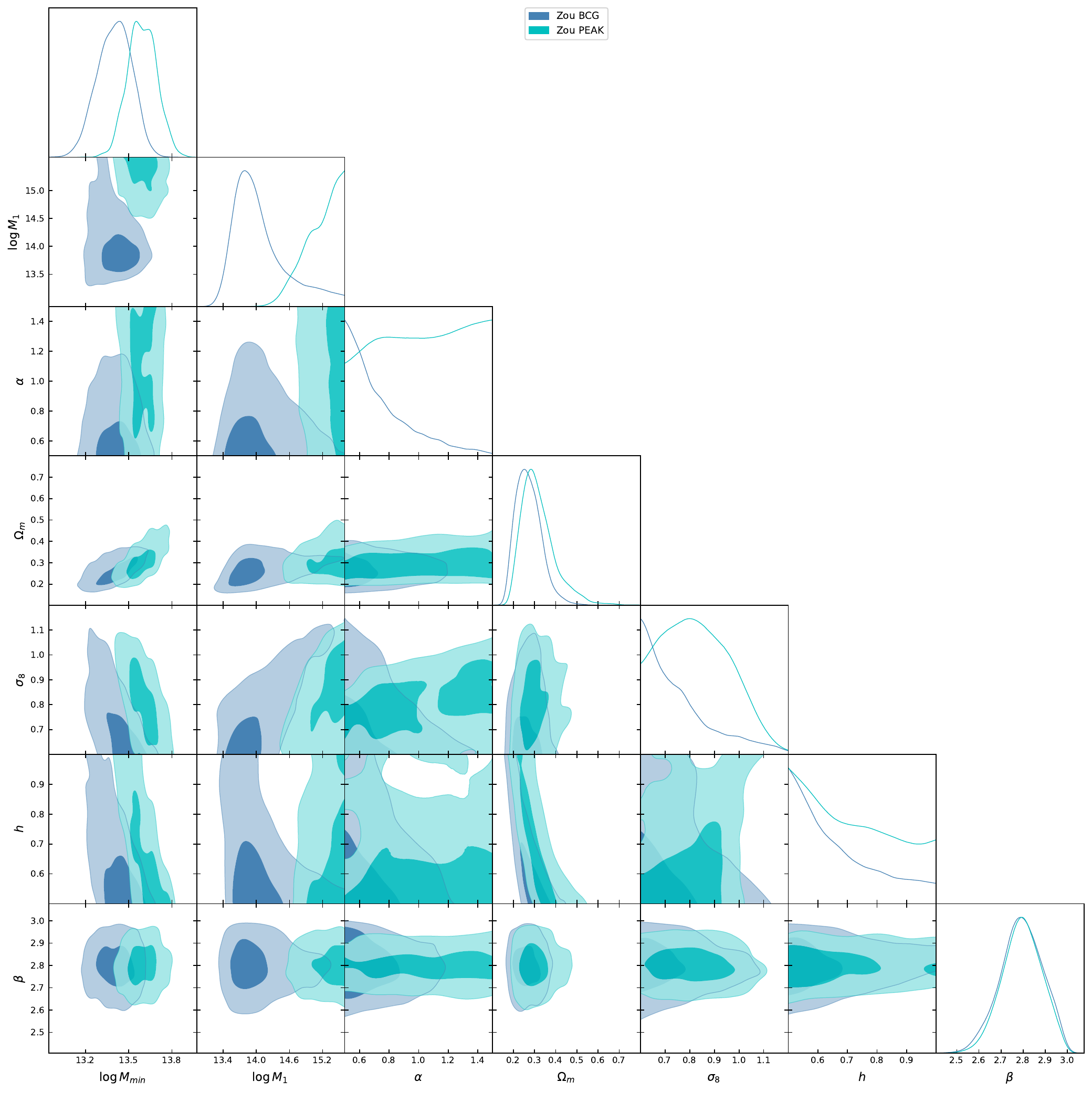}
  \caption{Marginalised posterior distributions and probability contours for the MCMC runs on the cross-correlation function using the two different available positions for ZOU cluster data. The dark blue fit was obtained using BCG positions, and the light blue fit was obtained using PEAK positions. In both cases, cross-correlation data above $\gtrsim 60$ arcmin were discarded.}
  \label{fig:zou_peak_corner}
\end{figure}

\begin{figure}[htbp]
  \centering
  \includegraphics[width=1\textwidth]{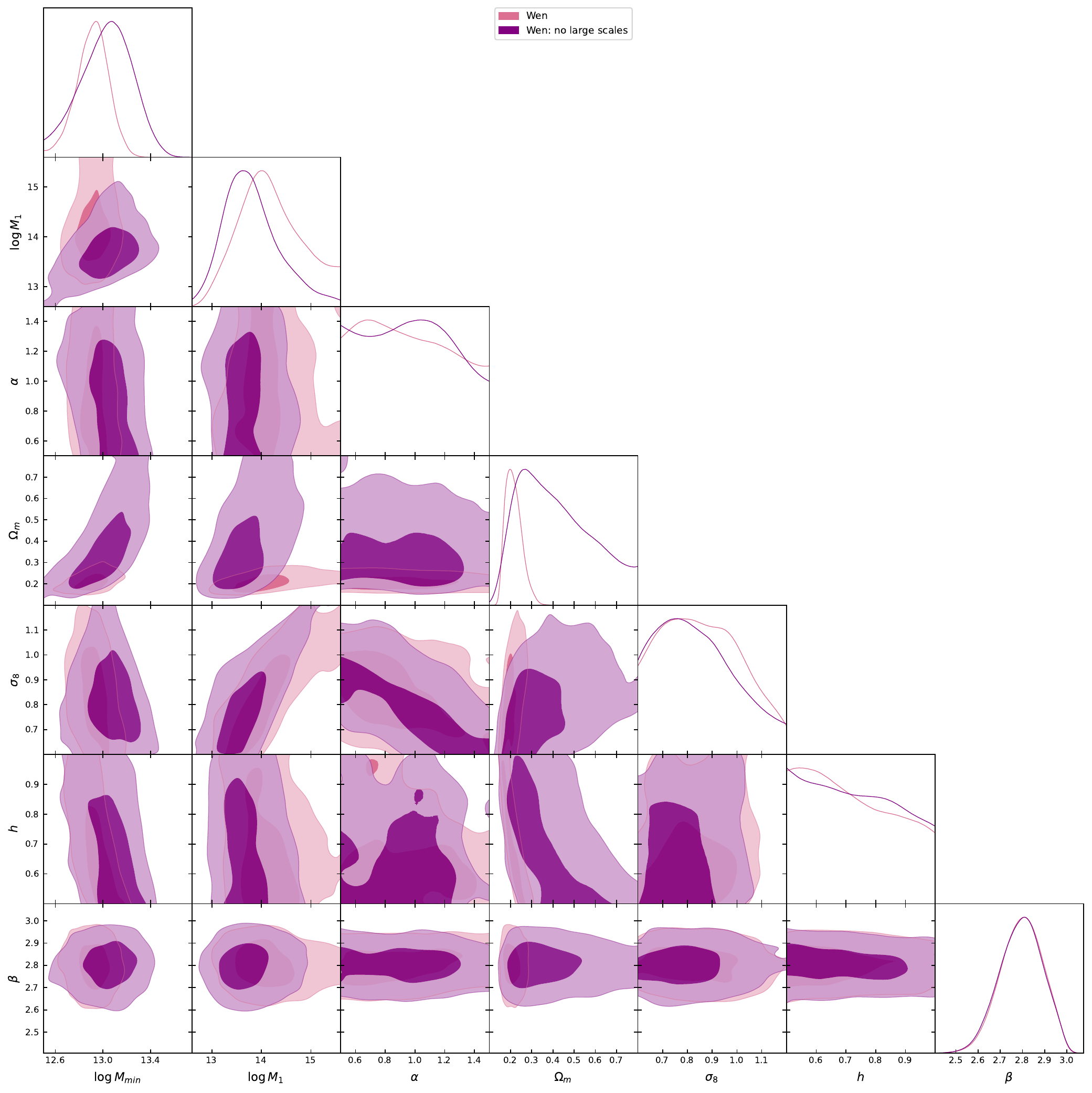}
  \caption{Marginalised posterior distributions and probability contours for the MCMC runs on the cross-correlation function using WEN cluster data. The pink fit was obtained using all data points. The violet fit was obtained excluding cross-correlation data above $\gtrsim 60$ arcmin.}
  \label{fig:wen_corner}
\end{figure}

Marginalised posterior distributions and probability contours of all the parameters involved in the MCMC runs discussed in this work. Figs. \ref{fig:Zou_full_corner} and \ref{fig:zou_peak_corner} depict the ZOU cases and Fig. \ref{fig:wen_corner} the WEN case.

\clearpage
\onecolumn
\section{Analysis of the cross-correlation function in individual GAMA zones}
\label{app:xcorr_zones}

This section presents the analysis of the cross-correlation function performed individually for each GAMA region.

\begin{figure*}[htp]
  \centering
  \includegraphics[width=1\textwidth]{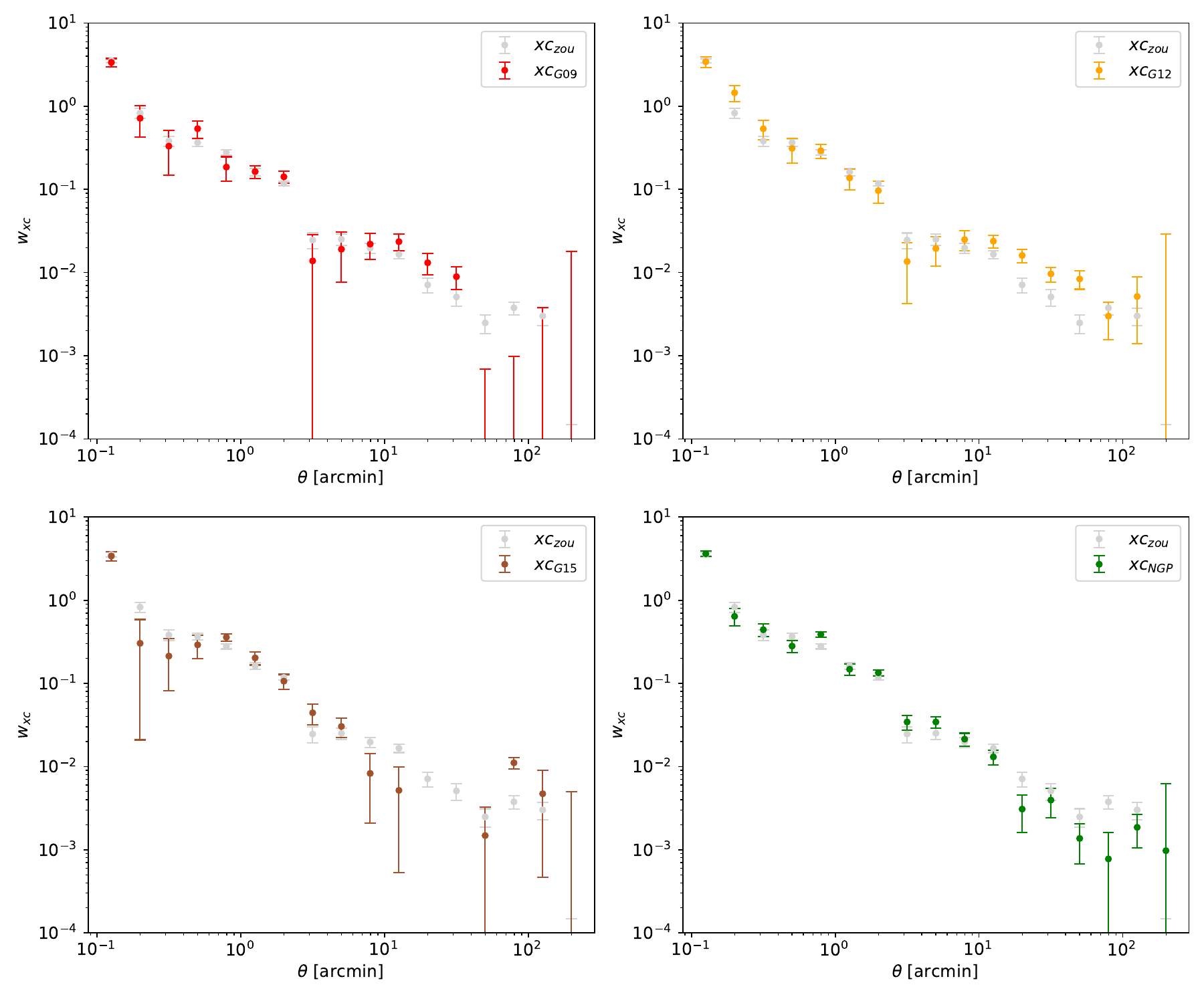}
  
  \caption{Cross-correlation data acquired for the ZOU sample was analysed across different zones. The panels depict G09 in red, G12 in yellow, G15 in brown, and NGP in green. The light grey data appearing in each panel correspond to the cross-correlation obtained using all four zones.
  }
  \label{fig:zou_zones}
\end{figure*}

Figure \ref{fig:zou_zones} shows the cross-correlation data collected for the ZOU sample. Notably, both the G12 and G15 exhibit a substantial increase in large-scale cross-correlation. In contrast, the G09 zone does not show such an increase, and the NGP zone displays minimal effects.

\begin{figure*}[htp]
  \centering
  \includegraphics[width=1\textwidth]{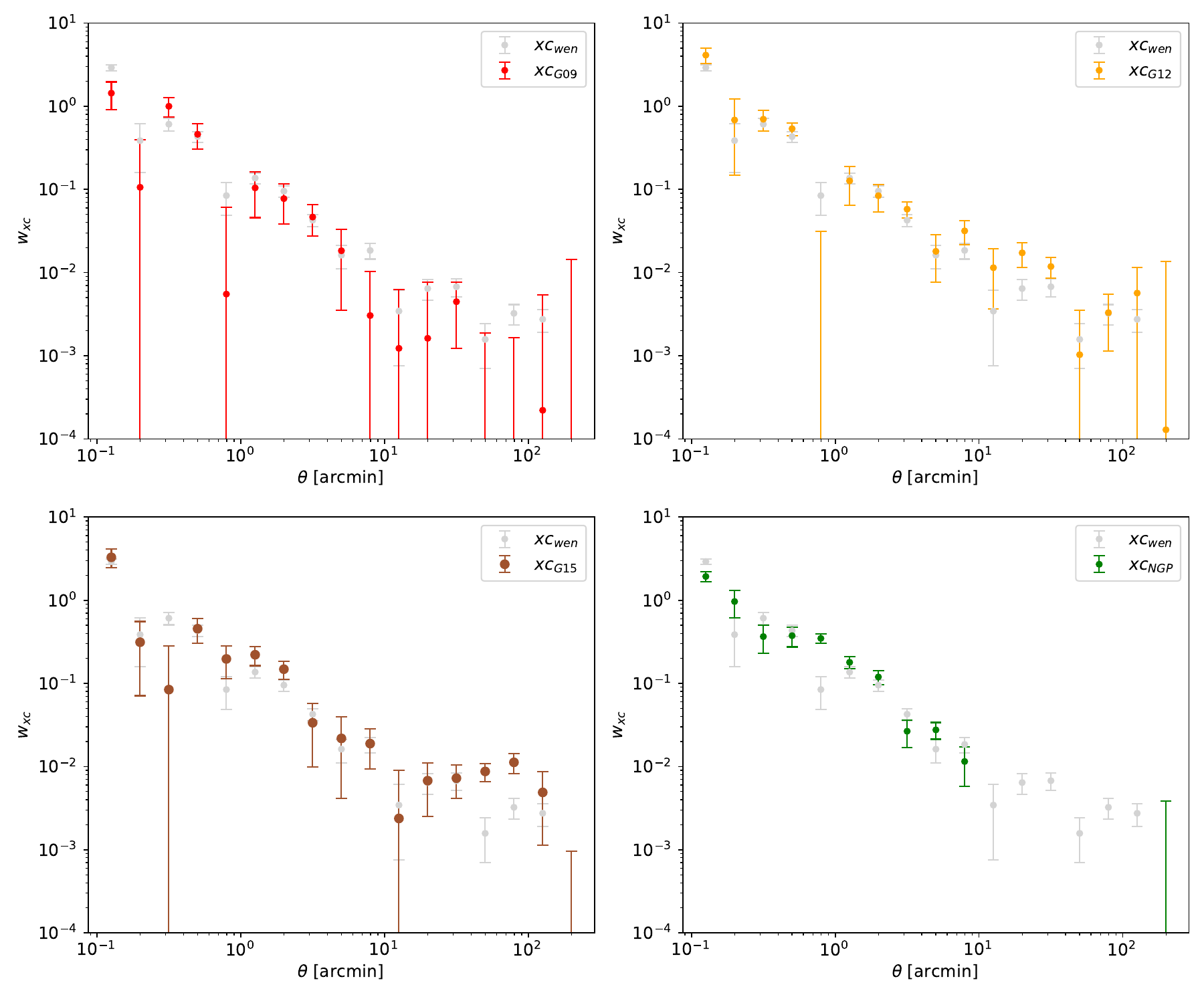}
 
  \caption{Cross-correlation data obtained for the WEN cluster analyzed across different zones. The panels depict G09 in red, G12 in yellow, G15 in brown, and NGP in green. The light gray data appearing in each panel correspond to the cross-correlation obtained using the entire cluster.}
  \label{fig:wen_zones}
\end{figure*}

Figure \ref{fig:wen_zones} depicts the cross-correlation data from the WEN cluster for the different individual zones. Similar to what was observed in Figure \ref{fig:zou_zones}, there is an excess of cross-correlation in G12 and G15 in this case as well. Conversely, there is a complete absence of this excess in NGP. It is worth noting that due to the lower number of sources in the WEN dataset, the results, when divided into zones, display significantly larger error bars.

\begin{figure*}[htp]
  \centering
  \includegraphics[width=1\textwidth]{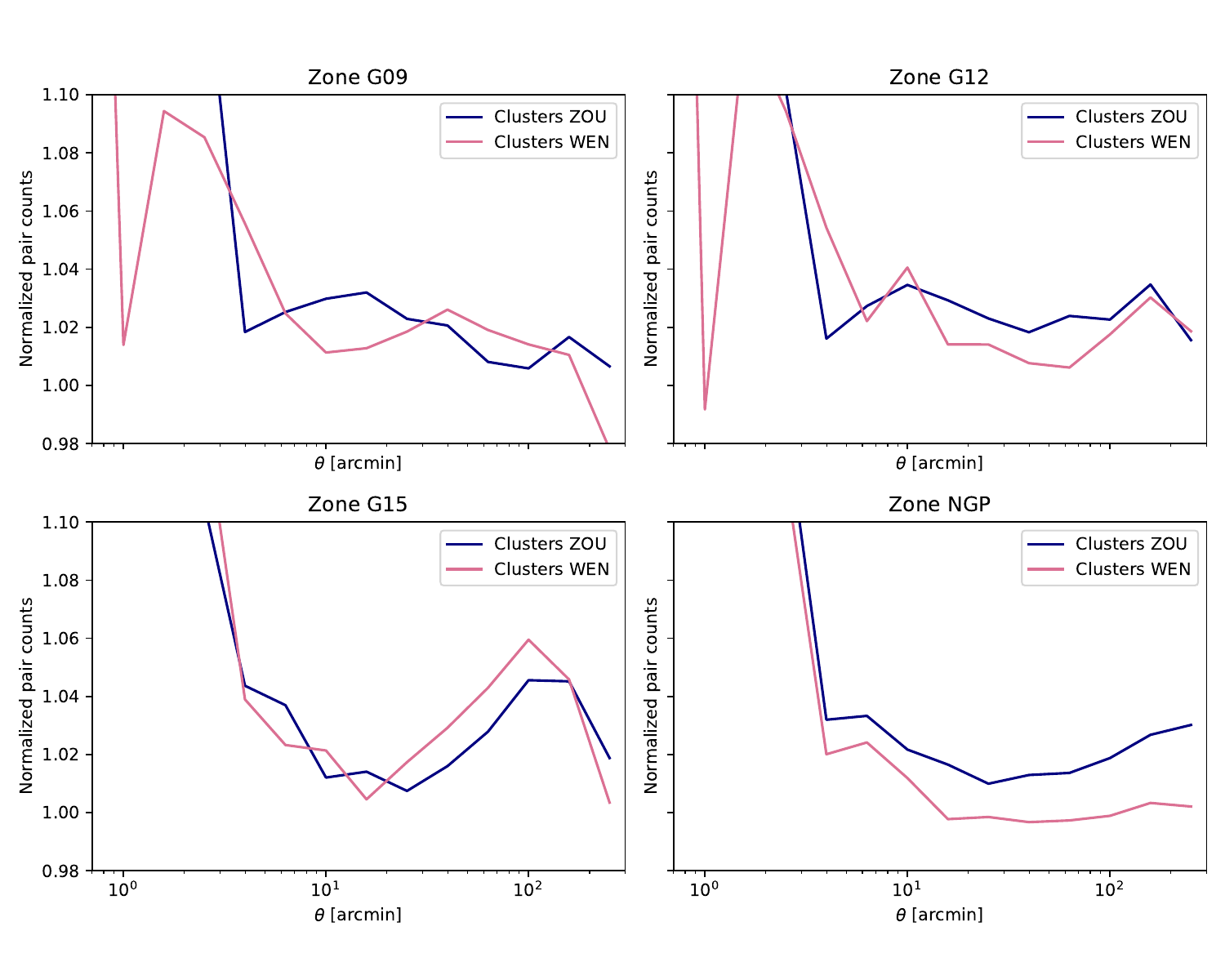}
  \caption{Foreground-background pair counts for each field obtained for the ZOU cluster (blue) and the WEN cluster (pink). Pair counts are normalised to those obtained between mock foreground and background random catalogues of the same sky regions. A high number of pairs is observed at small angular distances in every region, followed by a steep decline towards a value of 1, as expected in lensing. Regions G12, NGP, and particularly G15 exhibit an additional increase in pairs above 60 arcminutes, which does not fit the current lensing model. Region G09 does not show this anomalous behaviour.
  }
  \label{fig:npairs}
\end{figure*}

Figure \ref{fig:npairs} depicts the zone-wise foreground-background pair counts for both catalogues. The normalisation of pair counts is based on comparisons with mock foreground and background random catalogues corresponding to the same celestial areas. At the largest angular scales, the data-data pair counts are expected to converge to the random ones. Contrary to expectations, the G12, NGP and particularly the G15 zones show an increase in pair counts beyond 60 arcminutes, a phenomenon that deviates from the typical magnification bias behaviour. This observed anomaly, consistent across both cluster samples, is also observed in studies involving galaxy samples \citep{Cue23}. The absence of this surplus in the G09 zone accounts for the more discrete impact on the overall cross-correlation signal with respect to the one observed with galaxy samples.

\end{appendix}
\end{document}